\documentclass[11pt,english,aps,manuscript,aps,preprint,nofootinbib,,eqsecnum,superscriptaddress]{revtex4}
\usepackage[T1]{fontenc}
\usepackage[utf8]{luainputenc}
\usepackage{amsmath}
\usepackage{graphicx}
\usepackage{esint}
\usepackage{lipsum}

\makeatletter
\renewcommand\frontmatter@abstractwidth{0.4in}
\@ifundefined{textcolor}{}
{%
 \definecolor{BLACK}{gray}{0}
 \definecolor{WHITE}{gray}{1}
 \definecolor{RED}{rgb}{1,0,0}
 \definecolor{GREEN}{rgb}{0,1,0}
 \definecolor{BLUE}{rgb}{0,0,1}
 \definecolor{CYAN}{cmyk}{1,0,0,0}
 \definecolor{MAGENTA}{cmyk}{0,1,0,0}
 \definecolor{YELLOW}{cmyk}{0,0,1,0}
}
\numberwithin{figure}{section}

\usepackage[T1]{fontenc}
\usepackage[utf8]{luainputenc}
\usepackage{array}
\usepackage{appendix}
\usepackage{multirow}
\usepackage{amsmath}
\allowdisplaybreaks[4]
\usepackage{amssymb}
\usepackage{float}
\usepackage{wasysym}
\usepackage{graphicx}
\usepackage{esint}
\usepackage[marginal]{footmisc}
\usepackage{braket}
\usepackage[utf8]{inputenc}
\usepackage{verbatim}
\numberwithin{equation}{section}
\DeclareMathSizes{12}{12}{7}{7}

\renewcommand{\baselinestretch}{1.2}
\makeatother

\usepackage{babel}
\begin{document}

\title{The wavefunction reconstruction effects in calculation of DM-induced electronic
transition in semiconductor targets}

\author{Zheng-Liang Liang}
\email{liangzl@itp.ac.cn}
\affiliation{Institute of Applied Physics and Computational Mathematics\\Beijing, 100088, China}

\author{Lin Zhang}
\email{zhanglin@itp.ac.cn}
\affiliation{School of Physical Sciences, University of Chinese Academy of Sciences,\\Beijing, 100049, China}
\affiliation{CAS Key Laboratory of Theoretical Physics,\\Institute of Theoretical Physics, Chinese Academy of Sciences,\\Beijing, 100190, P.R. China}

\author{Ping Zhang}
\email{zhang_ping@iapcm.ac.cn}
\affiliation{Institute of Applied Physics and Computational Mathematics\\Beijing, 100088, China}

\author{Fawei Zheng}
\email{zheng_fawei@iapcm.ac.cn}
\affiliation{Institute of Applied Physics and Computational Mathematics\\Beijing, 100088, China}

\vspace{3cm}

\begin{abstract}

The physics of the electronic excitation in semiconductors induced
by sub-GeV dark matter (DM) have been extensively discussed in literature,
under the framework of the standard plane wave (PW) and pseudopotential
 calculation scheme. In this paper, we investigate the implication
of the all-electron (AE) reconstruction on  estimation of the DM-induced
electronic transition event rates. As a benchmark study, we first
calculate  the wavefunctions in silicon
and germanium bulk crystals based on
both the AE and pseudo (PS) schemes within the projector augmented
wave (PAW) framework, and then make comparisons between the  calculated
excitation event rates obtained from these two approaches. It turns out that in process
where large momentum transfer is kinetically allowed, the two calculated
event rates can differ by a factor of a few. Such discrepancies are
found to stem from the high-momentum components  neglected
in the PS scheme. It is thus implied  that the correction from the AE wavefunction in the core region is necessary for an accurate estimate of the DM-induced transition event rate in semiconductors.
\end{abstract}
\maketitle

\section{Introduction}

The existence of the dark matter (DM) has been well established on
both theoretical and phenomenological grounds in astrophysics and
cosmology. However, its nature still remains a mystery from a particle
physical perspective. Among various theoretical hypotheses, the generic
weakly interacting massive particles (WIMPs) have long been one of
the prevailing particle DM candidates. Their typical masses ranging
from GeV to TeV, and coupling strengths with the standard model (SM)
particles at the weak scale, not only naturally account for the observed
relic abundance in the context of thermal freeze-out, but also make
the WIMP direct detection possible for the present-day technologies~\cite{Goodman:1984dc}.
In the past three decades tremendous progress has been made in relevant
field and the sensitivity of the underground detectors will eventually
reach the solar neutrino background in the coming years, leaving only
a limited unexplored parameter region for the direct detection of
the WIMPs~\cite{Aprile:2016swn,Akerib:2016vxi,Tan:2016zwf}.

Facing this situation, both theorists and experimentalists begin to
shift their interest to other new directions. The sub-GeV DM as an
alternative consideration, with DM masses in the MeV to GeV range,
has attracted increasing attention in recent years, for its theoretical
motivations, and more importantly, its experimental feasibility. In
the sub-GeV DM paradigm, the DM particle is expected to reveal itself
by triggering ionization signals via the DM-electron scattering in
semiconductor detectors (\textit{e.g.}, the germanium-based detector EDELWEISS~\cite{Arnaud:2017usi} and SuperCDMS~\cite{Agnese:2014aze}  consisting of both silicon and germanium crystals) with energy thresholds as low as a few eV,
which translates to a DM mass detection threshold down to the MeV
scale. To realistically quantify the electronic excitation rate in
the crystal poses a challenge for theoretical estimation of relevant
experimental sensitivity, since the outer-shell electrons participating
in the excitation are delocalized in the crystal, and hence the isolated
atomic description turns invalid.

Step by step, several seminal studies help make concrete progress on
this issue~\cite{Essig:2011nj,Graham:2012su,Essig:2015cda,Lee:2015qva}.
In Refs.~\cite{Graham:2012su,Lee:2015qva}, where large momentum transfers
are assumed, the initial (valence) states are simply approximated
as corresponding free atomic orbitals, while the scattered ones are
described by plane waves. The calculations are performed with relevant
atomic \textit{Roothaan-Hartree-Fock} (RHF) wavefunctions in Ref.~\cite{Lee:2015qva}.
Another strategy is to calculate relevant electronic states under
the  framework of the \textit{density functional theory}
(DFT)~\cite{PhysRev.136.B864,PhysRev.140.A1133}, as adopted in Ref.~\cite{Essig:2015cda}, in which electronic
states are described with the band structures in the context of
solid state physics. The authors of Ref.~\cite{Essig:2015cda} adopted
software package $\mathtt{Quantum}\,\mathtt{Espresso}$~\cite{0953-8984-21-39-395502}
to perform the \textit{ab initio} pseudopotenial  calculations
on silicon and germanium semiconductors, hence putting the estimation
of electronic excitation rates onto a realistic ground, and encouraging
a recent wave of studies on a wider range of novel materials for the
detection of even lighter DM candidates~\cite{Hochberg:2016ajh,Hochberg:2016ntt,Hochberg:2016sqx,Hochberg:2017wce,Cavoto:2017otc,Knapen:2017ekk,Knapen:2017xzo,Griffin:2018bjn,Liang:2018wte}.

However, in this study our interest is focused on the effectiveness
of the pseudopotenial approaches adopted in the DFT implementation. While the periodic electronic wavefunctions are  conveniently expressed with  the \textit{plane wave} (PW) basis set in a typical
DFT computation of the electrons in crystalline solid, the rapidly varying wavefunction near the nucleus requires a huge number of PWs to be presented, even though they are irrelevant to most of the low-energy physical processes
(\textit{e.g.}, the chemical bonding properties). The purpose
of introducing  pseudopotenial is thus to soften the oscillatory core wavefunction and reduce the number of  PWs accordingly,  without disturbing the electrons outside the core region. Although the pseudization of the valence electrons has proved  an effective and efficient method in calculations of a wide range of electronic properties, the effectiveness
of such well-adopted procedure needs to be scrutinized in the study
of electronic excitation process, where transition energies are often deposited
through deep atomic inelastic scattering processes.  In particular, in
order to excite a valence electron far beyond the band gap, the incident
DM particle needs to penetrate into the core region where the pseudo
wavefunction description may no longer suffice for an accurate analysis. Consequently, a realistic description of inner electrons becomes necessary.
In fact, similar investigations have existed for long in literature,
for instance, exploring discrepancy between the \textit{pseudo} (PS) and the \textit{all-electron}
(AE) descriptions in optical properties of semiconductors~\cite{PhysRevB.63.125108},
and in X-ray Compton profiles for semiconductors and metals~\cite{PhysRevB.58.4320,MAKKONEN20051128}.
Following these literatures in which the AE  wavefunctions are restored from various reconstruction schemes, we use the term reconstruction effects
referring to  discrepancies in calculation originating from the PS  and AE  approaches.

For DM particles in the MeV-mass range, such investigations are even
more imperative, because compared to the case of optical excitation, due to different dispersion relations,
a much larger momentum transfer will be cost onto the massive particle
given the same amount of deposited energy in solids. However, high-momentum components of the PW expansion that would response to such large transferred momentum are systematically abandoned in the pseudopotenial calculations, and this may result in  a noticeable discrepancy between the two approaches. Common pseudopotenial-like
methods include \textit{norm-conserving} (NC)~\cite{PhysRevLett.43.1494},
\textit{ultrasoft} (US)~\cite{PhysRevB.41.7892}, and \textit{projector
augmented wave} (PAW)~\cite{PhysRevB.50.17953}, \textit{etc}..  For example, in calculation of the DM-induced transition event rates, the US  and PAW pseudopotenials were adopted in Ref.~\cite{Essig:2015cda} and Ref.~\cite{Hochberg:2017wce}, respectively.  As
a benchmark study, in this work we choose the PAW method as a concrete
example, since the PAW scheme not only well describes electrons in
the interstitial region, but also are capable of restoring the AE
wavefunctions in the core region. Besides, it is convenient to make
comparison within the same framework, and qualitative results can apply
to other pseudopotenial methods.

This paper is outlined as follows. We begin Sec.~\ref{sec:TransitionRate}
by giving a summary of concepts and formulae associated with the calculation
of the electronic transition rates induced by DM particles. In Sec.~\ref{sec:PAW}
we first take a brief review of the PAW method, and then generate
appropriate PAW datasets for the computation of the band structures.
By inputting these PAW files into the $\mathtt{ABINIT}$ package,
in Sec.~\ref{sec:ReconstructionEffects} we calculate the DM-induced
excitation events with both the AE and PW approaches for silicon and
germanium semiconductors, respectively. The ratios of the AE event
rates to the PW ones are calculated and the reconstruction effects
are explored. We conclude in Sec.~\ref{sec:Conclusion}.

 We use natural units $\hbar=c=1$ in all formulae in this paper, while velocities are expressed with units of $\mathrm{km\cdot s^{-1}}$ in text for convenience.

\section{\label{sec:TransitionRate}DM-induced electronic transition rates}

In this section, we give a summary of the formulae concerning the
DM-induced electronic transition event rates in the semiconductor
targets. For simplicity, in this work we only focus on the scenario
where the amplitude for DM particle scattering with a free electron
is constant, \textit{i.e.}, $\mathcal{M}_{\mathrm{free}}\left(\mathbf{q}\right)\propto1$,
and hence the DM-electron cross section $\sigma_{\chi e}=\left|\mathcal{M}_{\mathrm{free}}\right|^{2}\mu_{\chi e}^{2}/\pi$
is also a constant (with $\mu_{\chi e}=m_{\mathrm{e}}\, m_{\chi}/\left(m_{\mathrm{e}}+m_{\chi}\right)$
being the reduced mass of DM-electron pair), independent of DM transferred
momentum $\mathbf{q}$ and incident velocity $\mathbf{w}$. As a result,
the transition event rate for a DM particle with incident velocity
$\mathbf{w}$ to excite a bound electron from level $1$ to level
$2$ can be expressed in the following schematic form~\cite{Essig:2015cda}:
\begin{eqnarray}
\sigma_{1\rightarrow2}w & = & \frac{\sigma_{\chi e}}{\mu_{\chi e}^{2}}\int\frac{\mathrm{d}^{3}q}{4\pi}\,\delta\left(\frac{q^{2}}{2\, m_{\chi}}-\mathbf{q}\cdot\mathbf{w}+\Delta E_{1\rightarrow2}\right)\left|f_{1\rightarrow2}\left(\mathbf{q}\right)\right|^{2},
\end{eqnarray}
where the form factor
\begin{eqnarray}
f_{1\rightarrow2}\left(\mathbf{q}\right) & = & \int\mathrm{d}^{3}x\,\psi_{2}^{*}\left(\mathbf{x}\right)\psi_{1}\left(\mathbf{x}\right)e^{-i\mathbf{q}\cdot\mathbf{x}}\label{eq:form factor_0}
\end{eqnarray}
 encodes the excitation probability for the bound states in the electron
system, with $\psi_{1}\left(\mathbf{x}\right)$ and $\psi_{2}\left(\mathbf{x}\right)$
being the wave functions of the initial and final electron states,
respectively, and $\Delta E_{1\rightarrow2}$ denotes the relevant
energy difference. Then after taking into account the DM halo profile,
the excitation event rate from state $1$ to $2$ can be expressed
as
\begin{eqnarray}
\mathcal{R}_{1\rightarrow2} & = & \frac{\rho_{\chi}}{m_{\chi}}\left\langle \sigma_{1\rightarrow2}w\right\rangle \nonumber \\
 & = & \frac{\rho_{\chi}}{m_{\chi}}\left(\frac{\sigma_{\chi e}}{4\pi\mu_{\chi e}^{2}}\right)\int\mathrm{d}^{3}q\,\int\mathrm{d}^{3}w\, f_{\chi}\left(\mathbf{w};\,\hat{\mathbf{q}}\right)\,\delta\left(\frac{q^{2}}{2m_{\chi}}-\mathbf{q}\cdot\mathbf{w}+\Delta E_{1\rightarrow2}\right)\left|f_{1\rightarrow2}\left(\mathbf{q}\right)\right|^{2},\nonumber \\
\nonumber \\
 & = & \frac{\rho_{\chi}}{m_{\chi}}\left(\frac{\sigma_{\chi e}}{4\pi\mu_{\chi e}^{2}}\right)\int\mathrm{d}^{3}q\,\int\frac{g_{\chi}\left(\mathbf{w};\,\hat{\mathbf{q}}\right)}{q\, w}\,\mathrm{d}w\,\mathrm{d}\phi_{\mathbf{\hat{\mathbf{q}}}\mathbf{w}}\,\Theta\left[w-w_{\mathrm{min}}\left(q,\,\Delta E_{1\rightarrow2}\right)\right]\nonumber \\
 &  & \times\left|f_{1\rightarrow2}\left(\mathbf{q}\right)\right|^{2},\label{eq:R_12}
\end{eqnarray}
where the bracket $\left\langle \cdots\right\rangle $ denotes the
average over the DM velocity distribution, $\rho_{\chi}$ represents
the DM local density, $f_{\chi}\left(\mathbf{w},\,\hat{\mathbf{q}}\right)$
is the DM velocity distribution with unit vector $\hat{\mathbf{q}}$
as its zenith direction, $g_{\chi}\left(\mathbf{w};\,\hat{\mathbf{q}}\right)\equiv w^{2}f_{\chi}\left(\mathbf{w};\,\hat{\mathbf{q}}\right)$,
and $\Theta$ is the Heaviside step function. While $\phi_{\mathbf{\hat{\mathbf{q}}}\mathbf{w}}$
is the azimuthal angle of the spherical coordinate system $\left(\mathbf{\hat{\mathbf{q}}};\mathbf{w}\right)$,
the polar angle $\theta_{\hat{\mathbf{q}}\mathbf{w}}$ has integrated
out the delta function in above derivation. For the given $q$ and
$\Delta E_{1\rightarrow2}$, function $w_{\mathrm{min}}$ determines
the minimum kinetically accessible velocity for the transition:
\begin{eqnarray}
w_{\mathrm{min}}\left(q,\,\Delta E_{1\rightarrow2}\right) & = & \frac{q}{2\, m_{\chi}}+\frac{\Delta E_{1\rightarrow2}}{q}.\label{eq:wmin}
\end{eqnarray}
In practice, the velocity distribution can be approximated as a truncated
Maxwellian form in the galactic rest frame, \textit{i.e.}, $f_{\chi}\left(\mathbf{w},\,\hat{\mathbf{q}}\right)\propto\exp\left[-\left|\mathbf{w}+\mathbf{v}_{\mathrm{e}}\right|^{2}/v_{0}^{2}\right]\,\Theta\left(v_{\mathrm{esc}}-\left|\mathbf{w}+\mathbf{v}_{\mathrm{e}}\right|\right)$,
with the earth's velocity $v_{\mathrm{e}}=230\,\mathrm{km\cdot s}^{-1}$,
the dispersion velocity $v_{0}=220\,\mathrm{km\cdot s}^{-1}$ and
the galactic escape velocity $v_{\mathrm{esc}}=544\,\mathrm{km\cdot s}^{-1}$.

When applied to the periodic lattice of a crystalline solid, the energy
level can be labeled with discrete band index $i$, and crystal momentum
$\mathbf{k}$ defined within the first\textit{ Brillouin Zone} (BZ),
corresponding to the eigen wave functions of the \textit{Kohn-Sham}
(KS) orbitals. By taking advantage of the Bloch's theorem, these
wave functions can be written as a product of the phase factor $e^{i\mathbf{k}\cdot\mathbf{x}}$
and lattice periodic function $u_{i\mathbf{k}}\left(\mathbf{x}\right)$
as the following,
\begin{eqnarray}
\psi_{i\mathbf{k}}\left(\mathbf{x}\right) & = & \sqrt{\frac{\varOmega}{V}}\, e^{i\mathbf{k}\cdot\mathbf{x}}\, u_{i\mathbf{k}}\left(\mathbf{x}\right)\nonumber \\
 & = & \sum_{\mathbf{G}}C_{\mathbf{k}+\mathbf{G}}^{i}\frac{e^{i\left(\mathbf{k}+\mathbf{G}\right)\cdot\mathbf{x}}}{\sqrt{V}},\label{eq:PeriodWaveFunction}
\end{eqnarray}
 where $\varOmega$ and $V$ denote the volume of  the \textit{
unit cell} (UC) and of the whole crystal, respectively, and $u_{i\mathbf{k}}\left(\mathbf{x}\right)$
is represented as a linear combination of PWs  denoted by reciprocal
lattice vectors $\mathbf{G}$'s. For convenience, the periodic wave functions $\left\{ u_{i\mathbf{k}}\right\}$ are normalized within the UC, while $\left\{ \psi_{i\mathbf{k}}\right\}
 $ are normalized on the whole crystal. Thus, the square of the form factor
form factor Eq.~(\ref{eq:R_12}) can be explicitly written as
\begin{eqnarray}
\left|f_{i\mathbf{k}\rightarrow i'\mathbf{k}'}\left(\mathbf{q}\right)\right|^{2} & = & \left|\left\langle \psi_{i'\mathbf{k}'}\right|e^{-i\mathbf{q}\cdot\mathbf{x}}\left|\psi_{i\mathbf{k}}\right\rangle \right|^{2}\nonumber \\
 & = & \frac{\left(2\pi\right)^{3}}{V}\sum_{\mathbf{G}}\left|\left\langle \psi_{i'\mathbf{k}'}\right|e^{-i\mathbf{\left(\mathbf{k}-\mathbf{k}'+\mathbf{G}\right)}\cdot\mathbf{x}}\left|\psi_{i\mathbf{k}}\right\rangle \right|^{2}\,\delta^{3}\left(\mathbf{k}-\mathbf{k}'+\mathbf{G}-\mathbf{q}\right)\nonumber \\
 & = & \frac{\left(2\pi\right)^{3}}{V}\sum_{\mathbf{G}}\left|\int_{\Omega}\mathrm{d}^{3}x\, u_{i'\mathbf{k}'}^{*}\left(\mathbf{x}\right)\, e^{-i\mathbf{\mathbf{G}}\cdot\mathbf{x}}\, u_{i\mathbf{k}}\left(\mathbf{x}\right)\right|\,\delta^{3}\left(\mathbf{k}-\mathbf{k}'+\mathbf{G}-\mathbf{q}\right),\label{eq:square_FormFactor}
\end{eqnarray}
where the integral $\int_{\Omega}\mathrm{d}^{3}x\left(\ldots\right)$
is performed over the UC. Then, after inserting Eq.~(\ref{eq:square_FormFactor})
into Eq.~(\ref{eq:R_12}) and taking into account the two degenerate
spin states, the total excitation event rate can be expressed as
\begin{eqnarray}
\mathcal{R} & = & \frac{\rho_{\chi}}{m_{\chi}}\left(\frac{4\pi^{2}\sigma_{\chi e}}{\mu_{\chi e}^{2}}\right)V\sum_{i'}^{c}\sum_{i}^{v}\int_{\mathrm{BZ}}\frac{\mathrm{d}^{3}k'}{\left(2\pi\right)^{3}}\frac{\mathrm{d}^{3}k}{\left(2\pi\right)^{3}}\,\sum_{\mathbf{G}}\left\{ \int\frac{g_{\chi}\left(\mathbf{w},\,\hat{\mathbf{q}}\right)}{w}\,\mathrm{d}w\,\mathrm{d}\phi_{\mathbf{\hat{\mathbf{q}}}\mathbf{w}}\,\right.\nonumber \\
 &  & \times\left.\Theta\left[w-w_{\mathrm{min}}\left(q,\, E_{i'\mathbf{k}'}-E_{i\mathbf{k}}\right)\right]\right|_{\mathbf{q}=\mathbf{k}-\mathbf{k}'+\mathbf{G}}\,\left.\frac{\left|\int_{\Omega}\mathrm{d}^{3}x\, u_{i'\mathbf{k}'}^{*}\left(\mathbf{x}\right)\, e^{-i\mathbf{\mathbf{G}}\cdot\mathbf{x}}\, u_{i\mathbf{k}}\left(\mathbf{x}\right)\right|^{2}}{\left|\mathbf{k}-\mathbf{k}'+\mathbf{G}\right|}\right\} ,\nonumber \\
\label{eq:total ExcitationRate}
\end{eqnarray}
where the summations are over both the initial valence band states
and the final conduction states.

On occasions where only the bonding properties are of concern, the
KS wave functions are usually calculated with the pseudopotential
method to avoid invoking too many PWs for complete representation
of the oscillatory wave functions near the nucleus, while keeping
intact the description of the delocalized electrons outside the core
regions. At variance with those cases, however, the excitation rate
in Eq.~(\ref{eq:total ExcitationRate}) explicitly depends on the
true wave functions within the core region. Thus, it is tempting to
give a quantitative comparison between the excitation event rates
computed respectively from the PS and AE approaches. For this purpose,
in this work we choose the PAW method for the benchmark study. For
one thing, the PAW formalism is closely resemblant to the widely
used US pseudopotential method~\cite{PhysRevB.41.7892}, also
generating soft pseudopotential and giving an even more efficient
and reliable description of material electronic structure~\cite{PhysRevB.59.1758}.
For another, within the same PAW framework, the true AE wave functions
can be restored consistently from the built-in transformation. These
two reasons make the PAW framework a natural host for the comparison
between the PS and AE approaches concerning the calculations of the
excitation event rates.

\section{\label{sec:PAW}PAW method}

In this section we will take a brief review on the  PAW method proposed originally in Ref.~\cite{PhysRevB.50.17953}.
The purpose is to calculate relevant physical quantities through transforming
the AE wave function problem into a computationally
convenient PS one. To avoid confusion, it should be noted
that in the context the term "all electron" (AE) is only referred to
the true wave functions, in contrast to the pseudopotential method
that smoothens the valence wave functions near the nucleus, rather
than the scheme where all the core electrons are put on the equal
footing in solving the \textit{Kohn-Sham} (KS) equations. In fact,
the PAW scheme is also based on the frozen-core approximation. If necessary,  unfreezing of lower-lying core states is
straightforward and convenient in the PAW method~\cite{PhysRevB.59.1758}. Imagine
one expands the true AE KS single particle wave function $\ket{\Psi}$
in terms of a set of true atomic wave functions \{$\ket{\phi_{i}^{a}}$\}
inside an atom-specific augmentation sphere $\Omega^{a}$ centered
at the atom site $a$, and a smooth wave function $\ket{\tilde{\psi}}$
in the interstitial region, respectively, as the following,
\begin{eqnarray}
\ket{\Psi} & = & \begin{cases}
\ensuremath{\sum\limits _{i}}\ket{\phi_{i}^{a}}c_{i}, & \mathrm{for}\:\mathbf{r}\in\Omega^{a}\\
\ket{\tilde{\psi}}, & \mathrm{for}\:\mathbf{r}\notin\Omega^{a},
\end{cases}
\end{eqnarray}
where the AE partial waves \{$\ket{\phi_{i}^{a}}$\} are the eigen wavefunctions of the KS Schrödinger equation for an isolated atom,
and expansion coefficients $\left\{ \ket{c_{i}}\right\} $ along with
smooth wave function $\ket{\tilde{\psi}}$ are left to be determined.
The index $i$ stands for the main and the angular momentum quantum
numbers $\left(n,\,\ell,\, m\right)$. From \{$\ket{\phi_{i}^{a}}$\},
one can also generate a complete set of corresponding PS partial waves
\{$\ket{\tilde{\phi}_{i}^{a}}$\} and an accompanying set of projector
functions \{$\ket{\tilde{p}_{i}^{a}}$\} inside the augmentation region
$\Omega^{a}$. By construction they are required to fulfill the condition
\begin{eqnarray}
\ensuremath{\sum\limits _{i}}\ket{\tilde{\phi}_{i}^{a}}\bra{\tilde{p}_{i}^{a}} & = & 1\label{eq:completeRelation}
\end{eqnarray}
 over the augmentation region. Outside the augmentation sphere $\Omega^{a}$,
the auxiliary smooth partial waves \{$\ket{\tilde{\phi}_{i}^{a}}$\}
are constructed so that they are identical to  their AE counterparts
\{$\ket{\phi_{i}^{a}}$\}. Since the AE and PS partial waves coincide
on the augmentation sphere, it is natural to make a smooth continuation
of wave function $\ket{\tilde{\psi}}$ inwards with smooth PS partial
waves in the following manner,
\begin{eqnarray}
\ket{\tilde{\Psi}} & = & \begin{cases}
\ensuremath{\sum\limits _{i}}\ket{\tilde{\phi}_{i}^{a}}c_{i}, & \mathrm{for}\:\mathbf{r}\in\Omega^{a}\\
\ket{\tilde{\psi}}, & \mathrm{for}\:\mathbf{r}\notin\Omega^{a}.
\end{cases}
\end{eqnarray}
As a result, the initial AE wave function can be recast in terms of
the PS one through the transformation
\begin{eqnarray}
\ket{\Psi} & = & \hat{\mathcal{T}}\ket{\tilde{\Psi}}\nonumber \\
 & = & \left[1+\sum_{a,\, i}\left(\ket{\phi_{i}^{a}}-\ket{\tilde{\phi}_{i}^{a}}\right)\bra{\tilde{p}_{i}^{a}}\right]\ket{\tilde{\Psi}}\nonumber \\
 & = & \ket{\tilde{\Psi}}+\sum_{a,\, i}\left(\ket{\phi_{i}^{a}}-\ket{\tilde{\phi}_{i}^{a}}\right)\ensuremath{\braket{\tilde{p}_{i}^{a}|\tilde{\Psi}}}.\label{eq:ReconstructionRelation}
\end{eqnarray}
From this particular form, the effective Hamiltonian for the PS wave
function can be expressed accordingly as
\begin{eqnarray}
\hat{\widetilde{H}}_{\mathrm{PAW}} & = & \hat{\mathcal{T}}^{\dagger}\hat{H}_{\mathrm{KS}}\hat{\mathcal{T}},\label{eq:modifiedKS}
\end{eqnarray}
which in turn yields the transformed eigenvalue equation for $\ket{\tilde{\Psi}}$,
\begin{eqnarray}
\hat{\widetilde{H}}_{\mathrm{PAW}}\ket{\tilde{\Psi}} & = & \epsilon\,\hat{\mathcal{T}}^{\dagger}\hat{\mathcal{T}}\ket{\tilde{\Psi}},\label{eq:effectiveEigenEquation}
\end{eqnarray}
with the same eigenvalues as the original KS equation. Now it is the
PS wave function $\ket{\tilde{\Psi}}$ that appears in the modified
KS equation Eq.~(\ref{eq:effectiveEigenEquation}), which is expected
to significantly facilitate the computation because the strong oscillating
components have been separated out. Consequently, once the PS wave
function $\ket{\tilde{\Psi}}$ is obtained, its AE counterpart can
be easily reconstructed via Eq.~(\ref{eq:ReconstructionRelation}).
Practical applications show that two partial waves per orbital are
usually sufficient to obtain accurate results~\cite{PhysRevB.50.17953,PhysRevB.59.1758}.
Further details of the PAW method can be found in Refs.~\cite{PhysRevB.50.17953,PhysRevB.59.1758}.

Now we delve into the computational details of the generation of the
PAW datasets for two most common semiconductor targets, silicon and
germanium. In practice, we use the $\mathtt{atompaw}$ code~\cite{HOLZWARTH2001329}
to obtain the sets of atomic AE and PS partial waves, along with the
projectors in a procedure similar to the one originally suggested
by Blöchl~\cite{PhysRevB.50.17953}. Two partial waves are generated
for each angular momentum quantum number $\ell$, and the GGA-PBE
exchange-correlation functional~\cite{PhysRevLett.77.3865} is chosen
for the calculation. The atomic AE partial wave is expressed in the
spherical coordinates as a product of a radial function times a spherical
harmonic function:
\begin{equation}
\ensuremath{\braket{\mathbf{r}|\phi_{i}^{a}}}\equiv\ensuremath{\braket{\mathbf{r}|\phi_{n\ell m}^{a}}}\equiv\psi_{n\ell}^{a}\left(r\right)Y_{\ell}^{m}\left(\hat{\mathbf{r}}\right)\equiv\frac{\phi_{n\ell}^{a}\left(r\right)}{r}Y_{\ell}^{m}\left(\hat{\mathbf{r}}\right),
\end{equation}
from which the radial KS Schrödinger eigen equation for the $\ell$-th
partial wave can be written as
\begin{equation}
\left[\frac{-1}{2\, m_{\mathrm{e}}}\left(\frac{\mathrm{d}^{2}}{\mathrm{d}r^{2}}+\frac{2}{r}\frac{\mathrm{d}}{\mathrm{d}r}-\frac{\ell\left(\ell+1\right)}{r^{2}}\right)+v_{\mathrm{KS}}^{\mathrm{at}}\left[n\right]\left(r\right)\right]\psi_{n\ell}\left(r\right)=\epsilon_{n\ell}\,\psi_{n\ell}\left(r\right),\label{eq:KSradialEquation}
\end{equation}
or
\begin{eqnarray}
\left[\frac{-1}{2\, m_{\mathrm{e}}}\left(\frac{\mathrm{d}^{2}}{\mathrm{d}r^{2}}-\frac{\ell\left(\ell+1\right)}{r^{2}}\right)+v_{\mathrm{KS}}^{\mathrm{at}}\left[n\right]\left(r\right)\right]\phi_{n\ell}\left(r\right) & = & \epsilon_{n\ell}\,\phi_{n\ell}\left(r\right),
\end{eqnarray}
where $v_{\mathrm{KS}}^{\mathrm{at}}\left[n\right]\left(r\right)$
is self-consistent atomic potential and $\epsilon_{n\ell}$ is the
energy eigenvalue. Based on these AE wave functions, corresponding
radial PS partial waves $\tilde{\phi}_{n\ell}\left(r\right)$ and
projector functions $p_{n\ell}\left(r\right)$ can then be constructed.
Relevant results for silicon and germanium are summarized below.

\subsection{Si}

\begin{figure}
\begin{centering}
\includegraphics[scale=0.75]{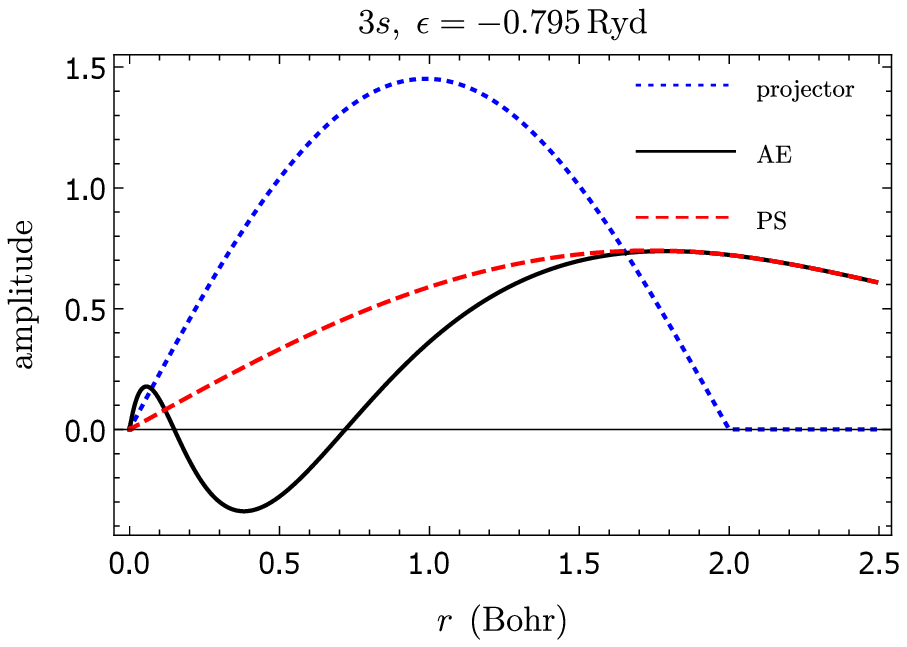}$\qquad$\includegraphics[scale=0.75]{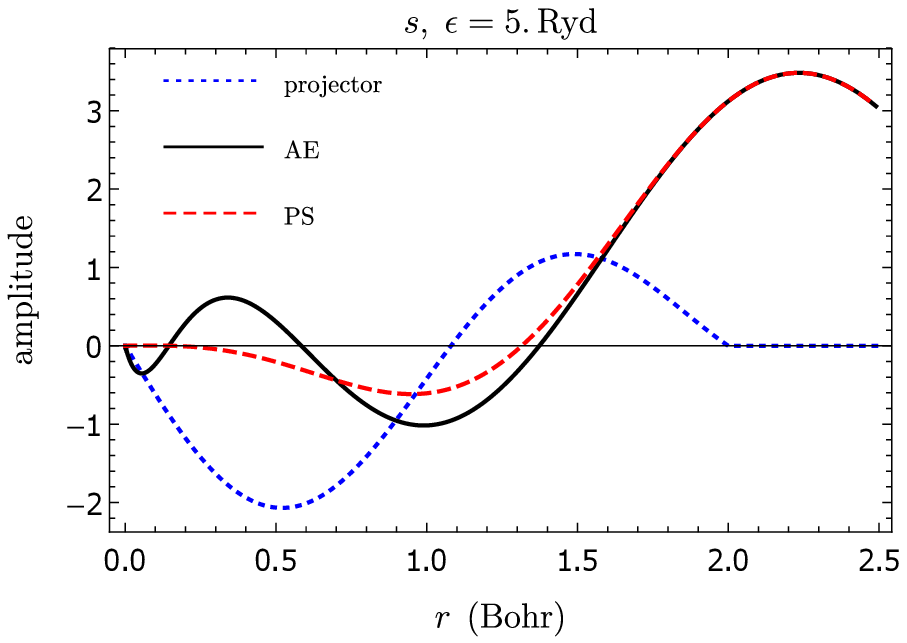}\vspace{0.3cm}

\par\end{centering}

\begin{centering}
\includegraphics[scale=0.75]{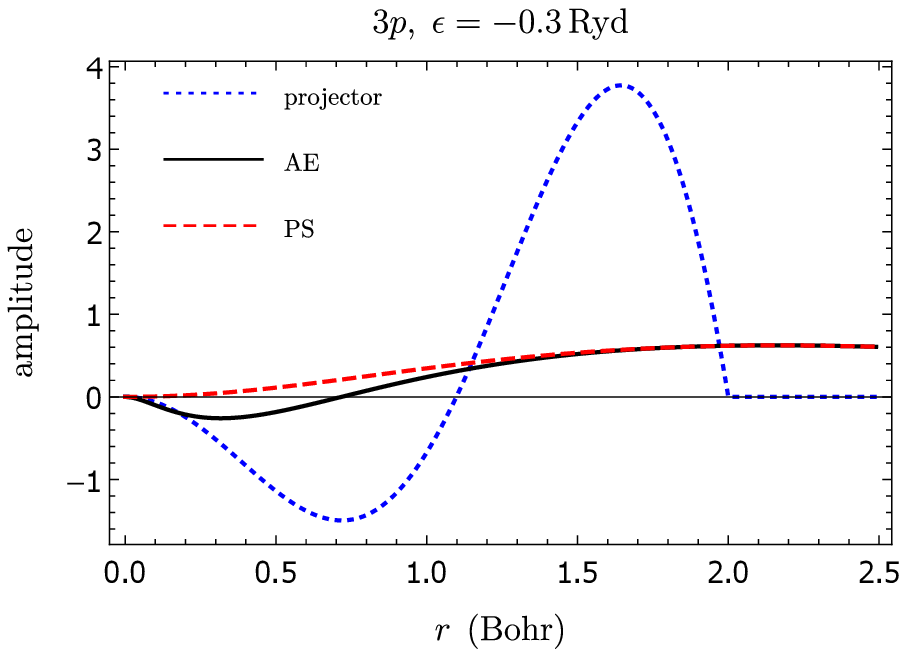}$\qquad$\includegraphics[scale=0.75]{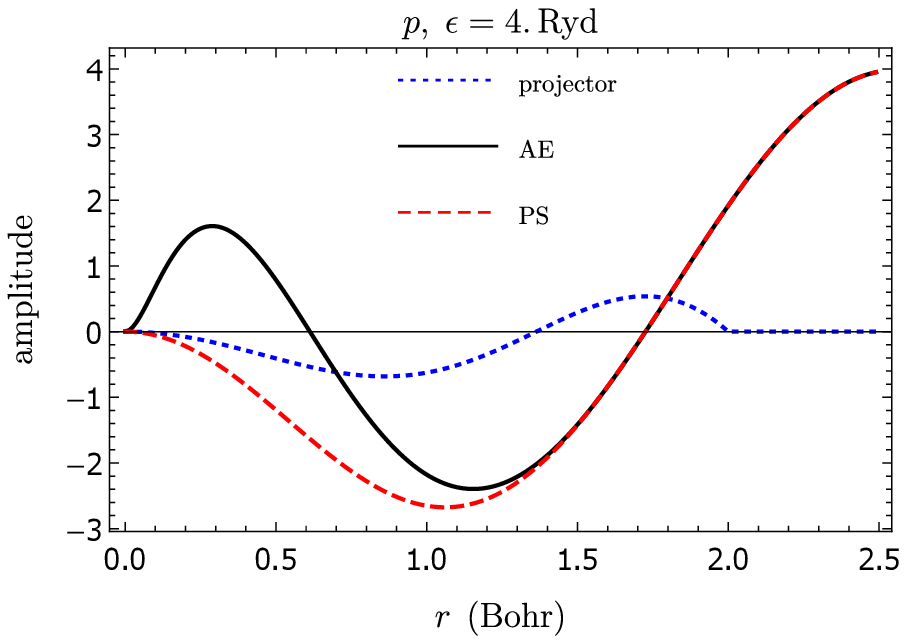}
\par\end{centering}

\vspace{0.3cm}

\begin{centering}
\includegraphics[scale=0.75]{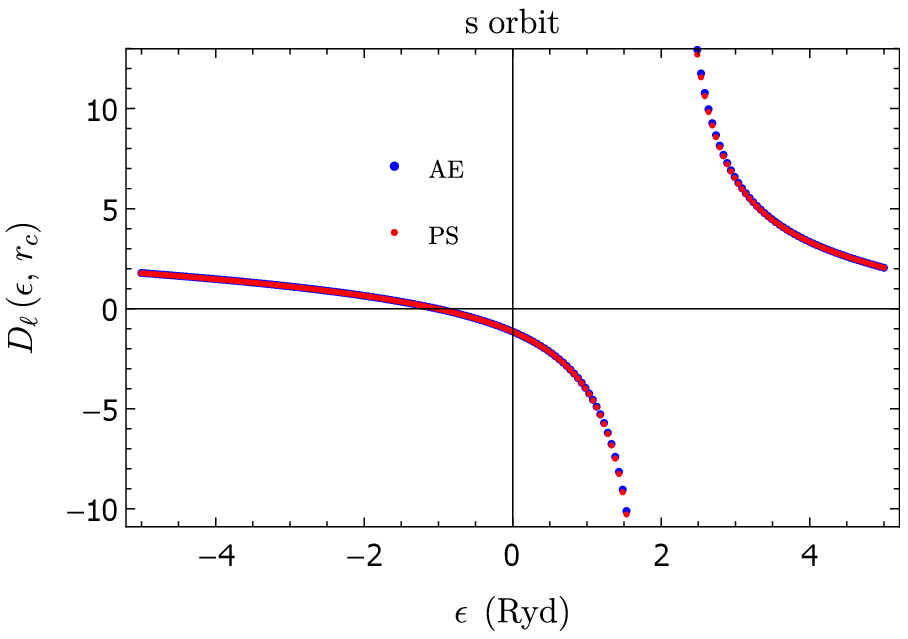}$\qquad$\includegraphics[scale=0.75]{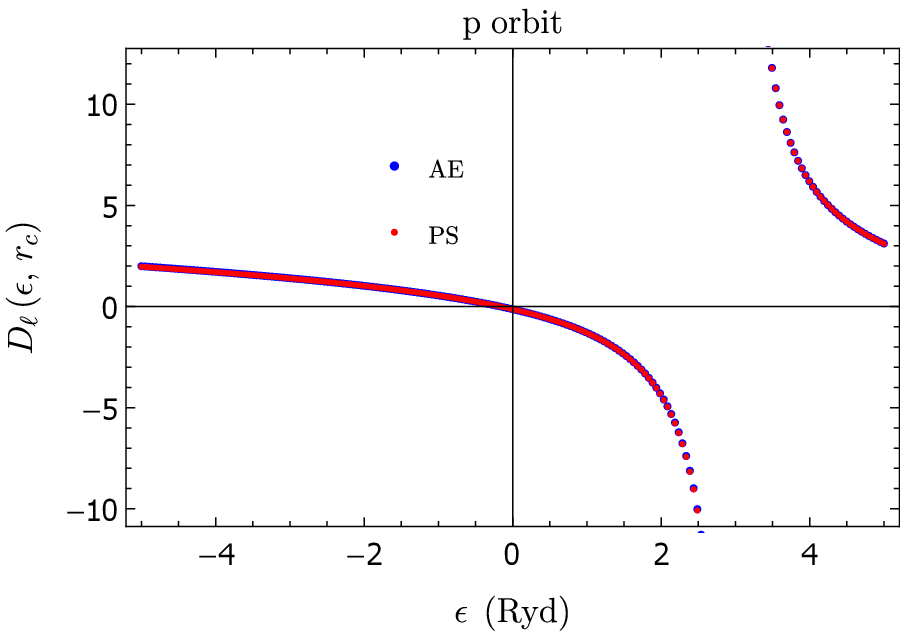}
\par\end{centering}

\protect\caption{\label{fig:Si_Pwaves} Radial partial waves and energy derivatives
for silicon. Shown in the first (second) row are two sets of AE (\textit{black
solid}), PS (\textit{red dashed}) partial waves and projectors (\textit{blue dotted})
for angular momentum $\ell=0\,\left(1\right)$, respectively, and
in the third row the energy derivatives of the AE (\textit{blue}) and PS (\textit{red})
partial waves for $\ell=0\,\left(1\right)$ are also presented. See
text for details.}
\end{figure}

\begin{figure}
\begin{centering}
\includegraphics[scale=0.75]{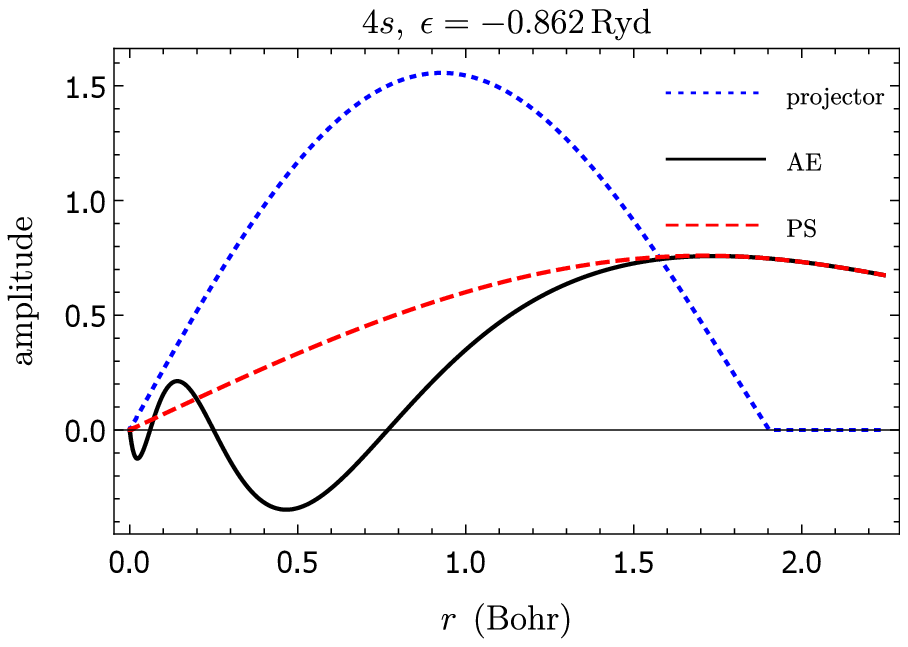}$\qquad$\includegraphics[scale=0.75]{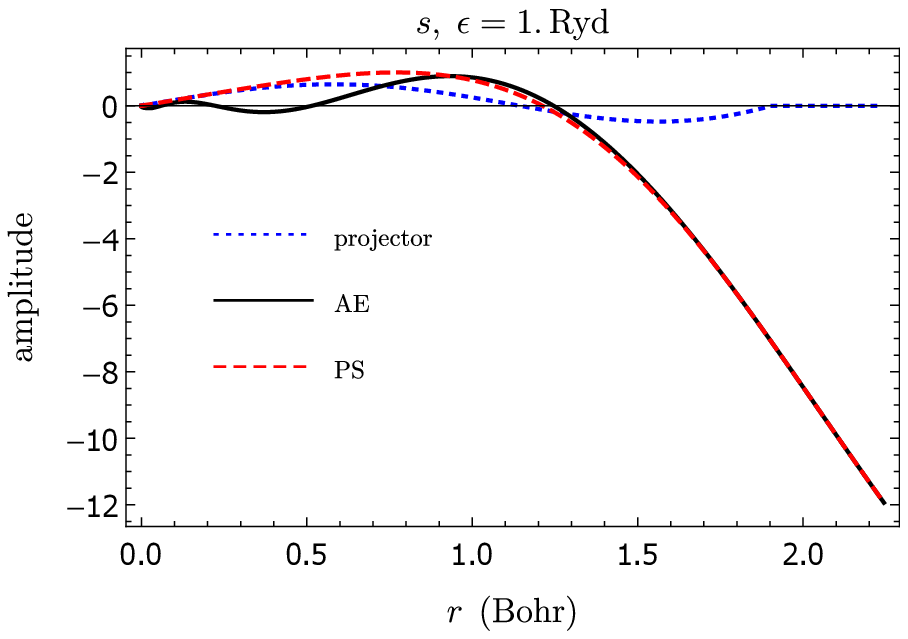}\vspace{0.3cm}

\par\end{centering}

\begin{centering}
\includegraphics[scale=0.75]{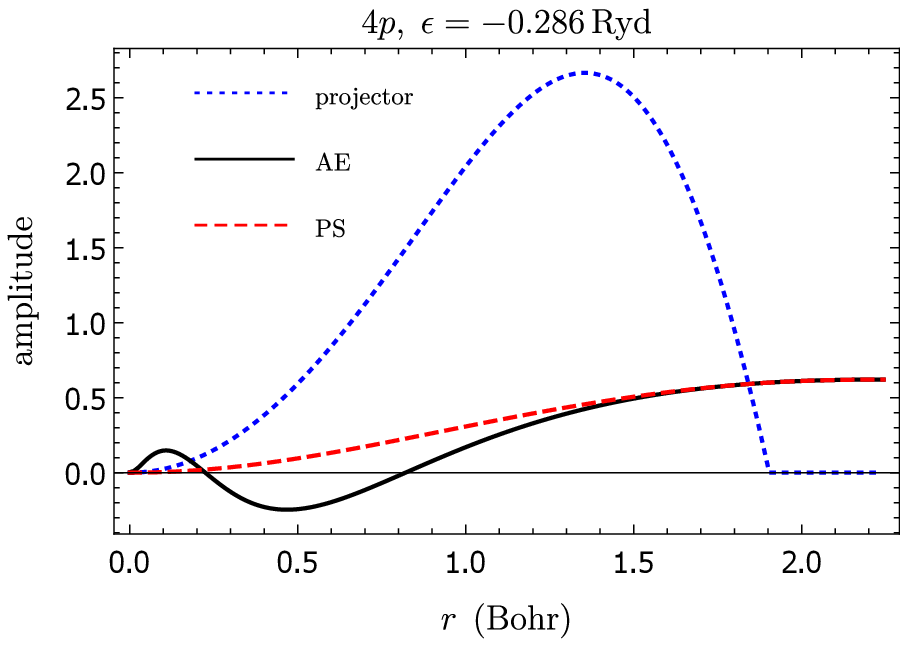}$\qquad$\includegraphics[scale=0.75]{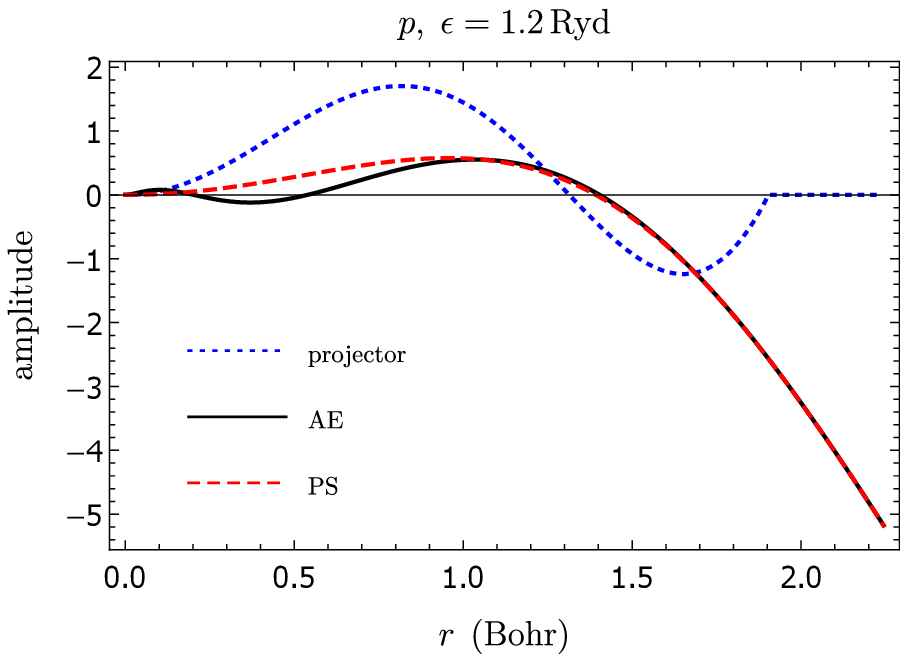}
\par\end{centering}

\vspace{0.3cm}

\begin{centering}
\includegraphics[scale=0.75]{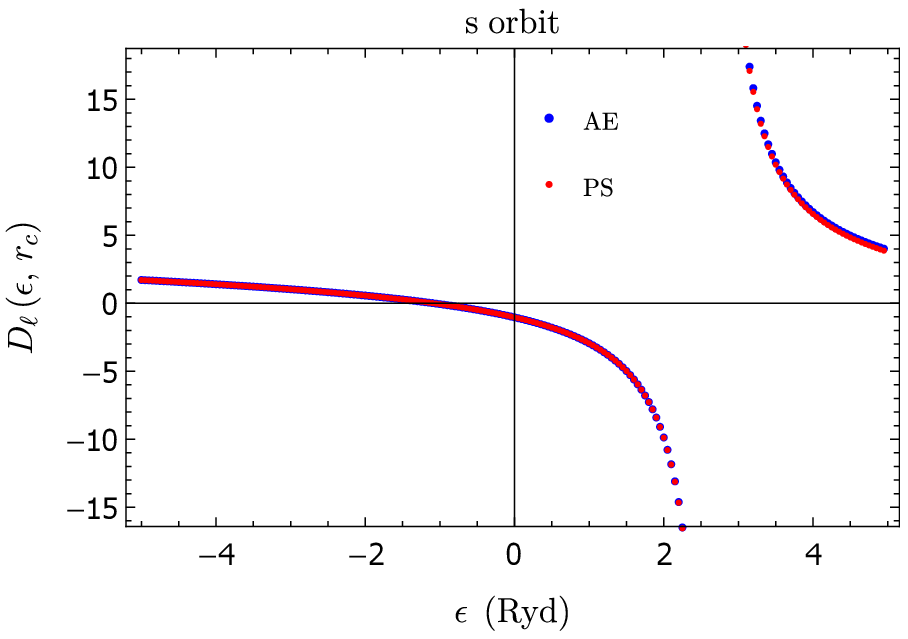}$\qquad$\includegraphics[scale=0.75]{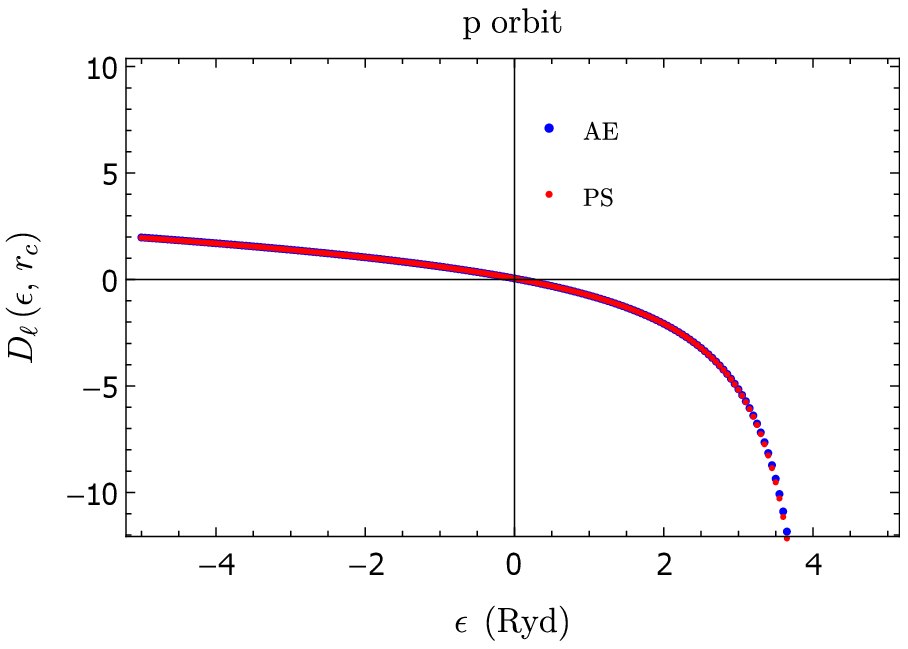}
\par\end{centering}

\protect\caption{\label{fig:Ge_Pwaves}Radial partial waves and energy derivatives
for germanium. Shown in the first (second) row are two sets of AE
(\textit{black solid}), PS (\textit{red dashed}) partial waves and projectors (\textit{blue
dotted}) for angular momentum $\ell=0\,\left(1\right)$, respectively,
and in the third row the energy derivatives of the AE (\textit{blue}) and PS
(\textit{red}) partial waves for $\ell=0\,\left(1\right)$ are also presented.
See text for details.}
\end{figure}

In generation of the PAW datasets of the silicon, the $3s$ and $3p$
electrons are taken as the valence electrons, while others in the
closed shell are approximated as the frozen core. In Fig.~\ref{fig:Si_Pwaves}
we present the partial waves and energy derivatives for the silicon
atom, following the $\mathtt{atompaw}$ recipe. Specifically, we adopt
a radius of the augmentation sphere $r_{c}=2.\,\mathrm{Bohr}$, and
two partial waves (so as the projectors) for each angular momentum
quantum number $\ell=0$ and $\ell=1$ are generated, based on two reference
energies of a bound and a scattering states. For $\ell=0\,\left(1\right)$,
the reference energies are respectively the eigen energies of the
bound state $\epsilon_{3s}=-0.795\,\mathrm{Ry}$ ($\epsilon_{3p}=-0.300\,\mathrm{Ry}$)
and the scattering state of the silicon atom $\epsilon_{s}=5.\,\mathrm{Ry}$
($\epsilon_{p}=4.\,\mathrm{Ry}$). In third row shown are the dimensionless
logarithmic derivatives $D_{\ell}$ for corresponding AE and PS atomic
partial waves, which is defined as
\begin{eqnarray}
D_{\ell}\left(\epsilon,\, r\right) & \equiv & r\,\frac{\partial\psi_{\ell}\left(\epsilon,r\right)}{\mathrm{\partial}r}/\psi_{\ell}\left(\epsilon,r\right),
\end{eqnarray}
where $\psi_{\ell}\left(\epsilon,r\right)$ is either the AE partial
wave function of Eq.~(\ref{eq:KSradialEquation}) at an arbitrary
reference energy $\epsilon$, or the PS counterpart from the transformed
version Eq.~(\ref{eq:effectiveEigenEquation}). For large enough
augmentation radius $r_{c}$, the quantity $D_{\ell}\left(\epsilon,\, r_{c}\right)$
uniquely determines the phase shifts of the partial waves in scattering
theory. Thus the coincidence of the logarithmic derivatives implies
that the constructed PAW Hamiltonian well recovers the scattering
properties of a single atom, indicating a good performance of the
PAW implementation in the energy range of interest.

\subsection{Ge}

In a parallel fashion, the corresponding partial waves and energy
derivatives for the germanium atom are presented in Fig.~\ref{fig:Ge_Pwaves}.
Similarly to case of silicon, the $4s$ and $4p$ orbitals are treated
as the valence states, while the closed shell is approximated as the
frozen core. With radius of the augmentation sphere $r_{c}=1.9\,\mathrm{Bohr}$,
two reference energies of the bound orbital $\epsilon_{4s}=-0.862\,\mathrm{Ry}$
($\epsilon_{4p}=-0.286\,\mathrm{Ry}$) and unbound state $\epsilon_{s}=1.\,\mathrm{Ry}$
($\epsilon_{p}=1.2\,\mathrm{Ry}$) are chosen to generate the $\ell=0\,\left(1\right)$
partial waves and projectors. For germanium, it is also shown that the logarithmic
derivatives of the AE and PS atomic partial waves coincide with each
other quite closely.

\section{\label{sec:ReconstructionEffects}Reconstruction effects on transition
event rate}

In this section, we first utilize the PAW datasets generated from
the $\mathtt{atompaw}$ code to calculate the band structures of silicon
and germanium, as well as both the corresponding AE and PS eigen wave
functions with the $\mathtt{ABINIT}$ package~\cite{GONZE20092582},
from which we then analyze and compare the calculations of relevant
DM-induced transition event rates.
\begin{figure}
\begin{centering}
\includegraphics[scale=0.6]{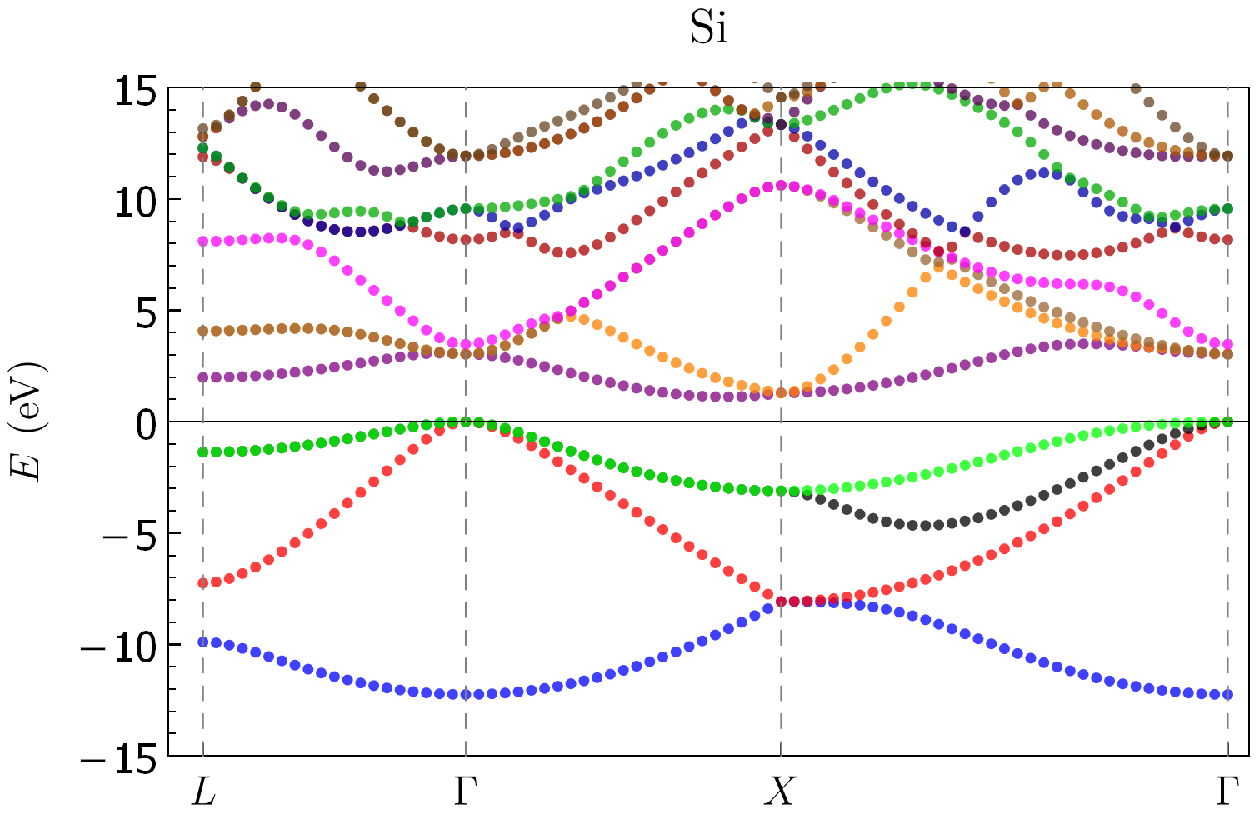}$\quad$\includegraphics[scale=0.6]{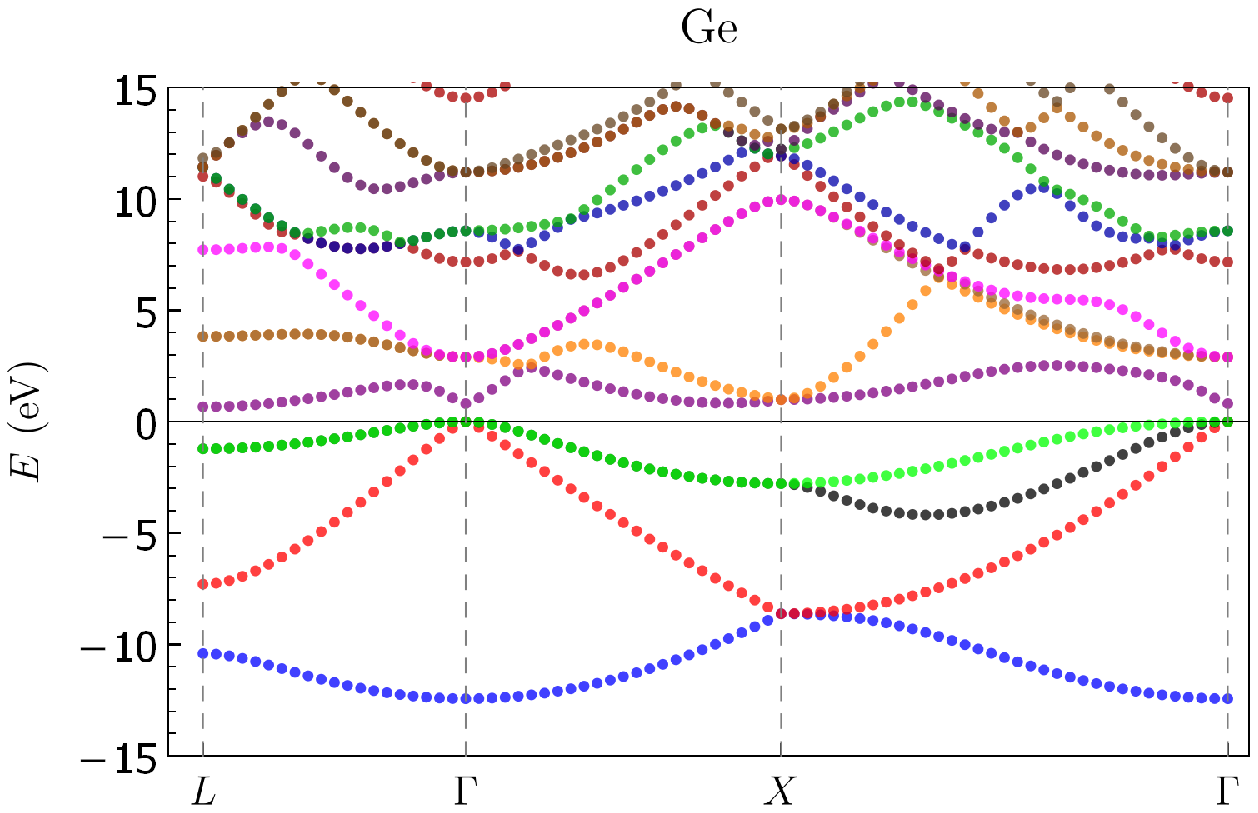}
\par\end{centering}

\protect\caption{\label{fig:band structure}The scissor corrected band structures of
silicon (\textit{left}) and germanium (\textit{right}) along the Brillouin
zone path $L$-$\Gamma$-$X$-$\Gamma$, respectively. The calculations
are performed using the $\mathtt{ABINIT}$ package~\cite{GONZE20092582},
based on the PAW data sets output from the $\mathtt{atompaw}$ code~\cite{HOLZWARTH2001329}.
The top of the valence bands are rescaled to $0\,\mathrm{eV}$, while
the dotted lines above the horizontal lines represent eleven lowest
conduction bands. See text for details.}
\end{figure}

\begin{figure}
\begin{centering}
\includegraphics[scale=0.2]{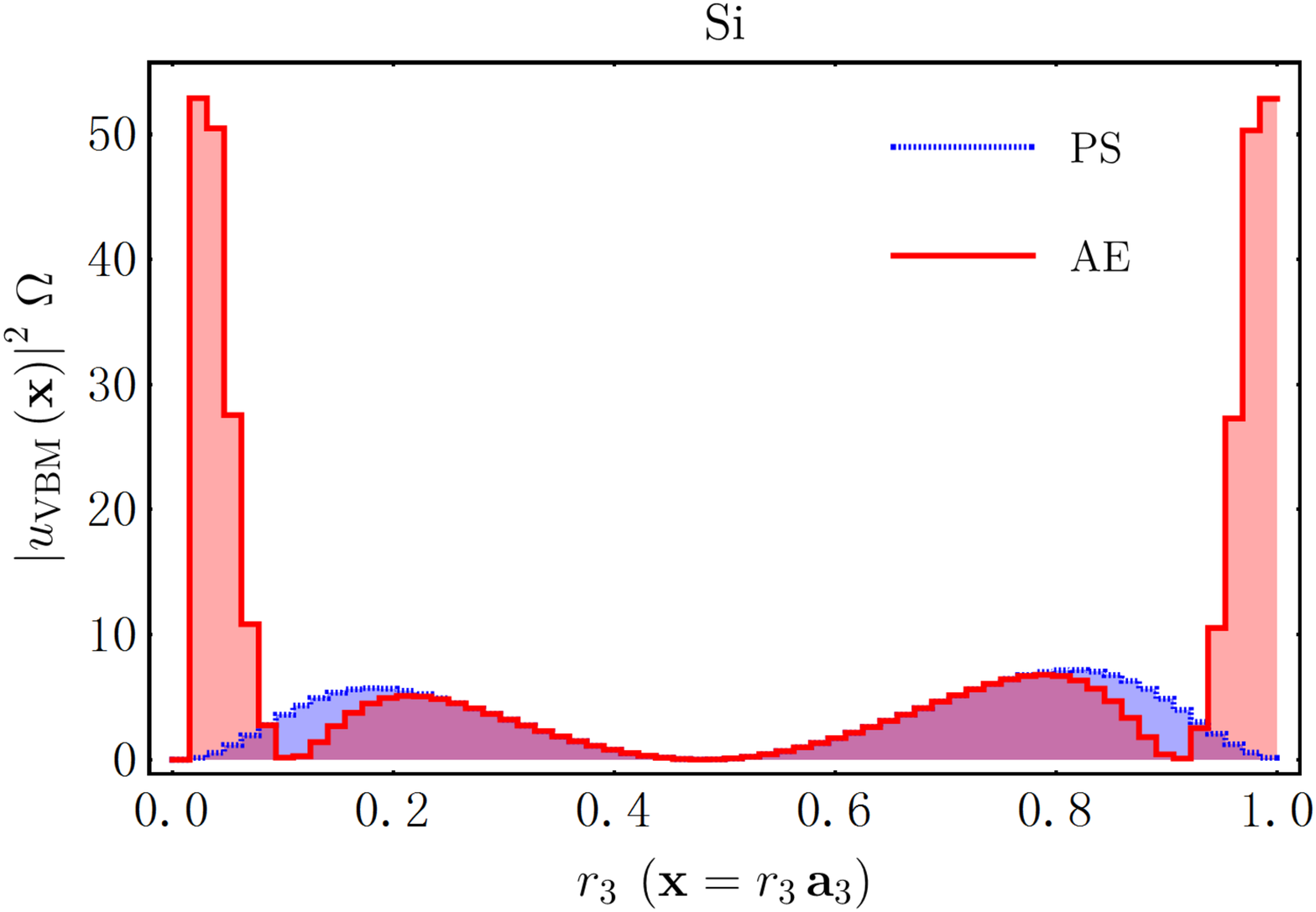}$\qquad$\includegraphics[scale=0.2]{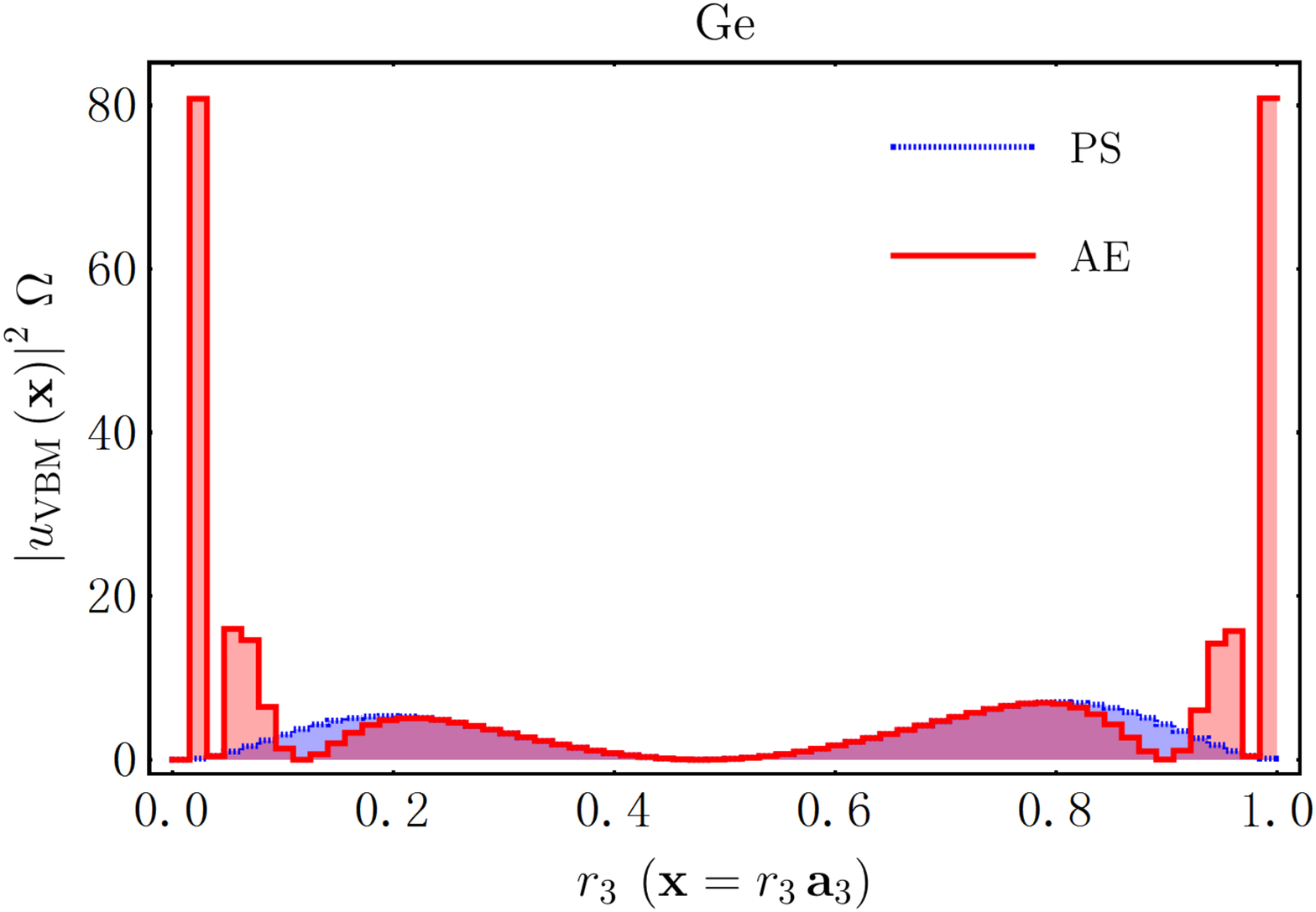}
\par\end{centering}

\begin{centering}
\includegraphics[scale=0.2]{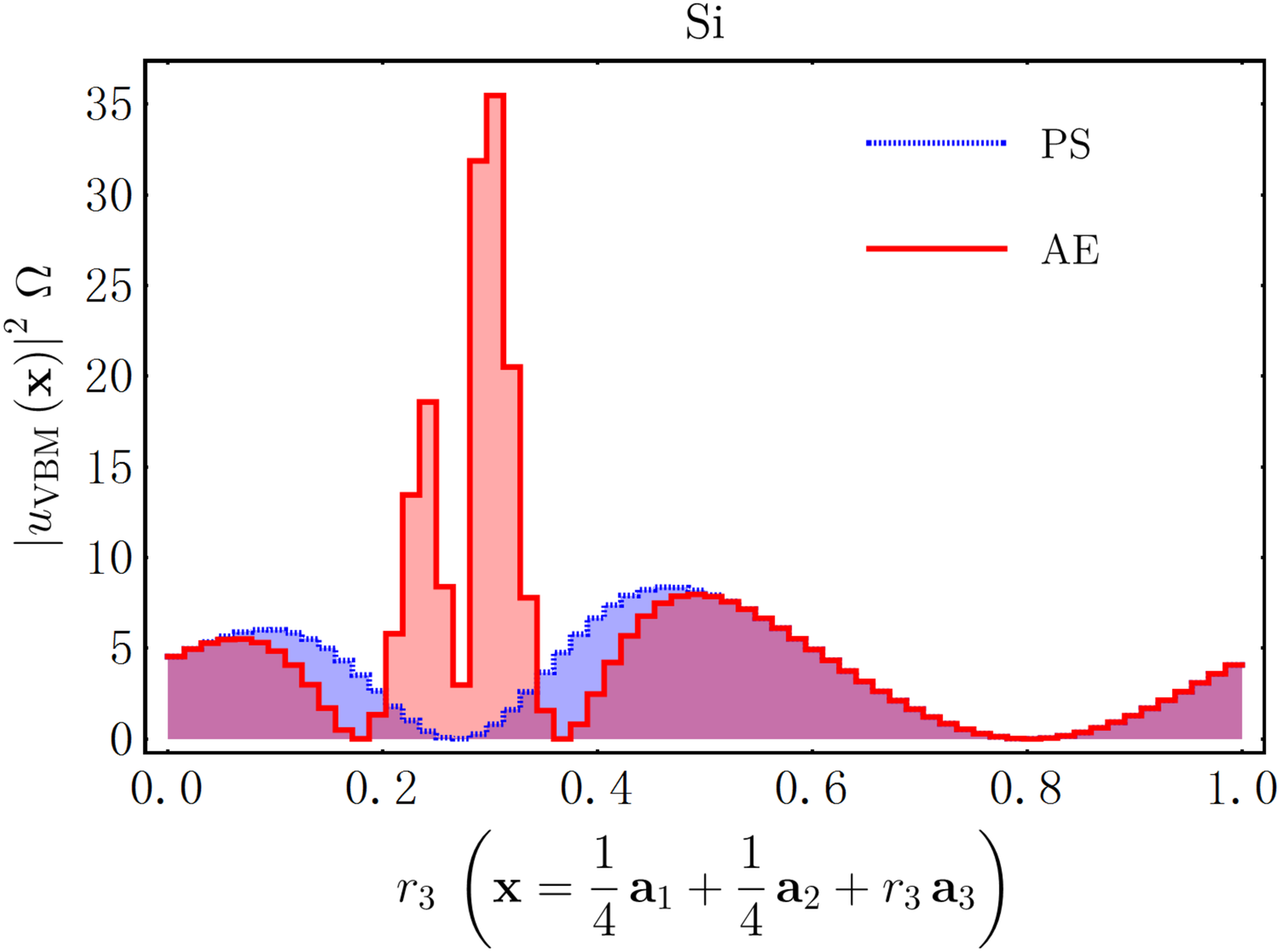}$\qquad$\includegraphics[scale=0.2]{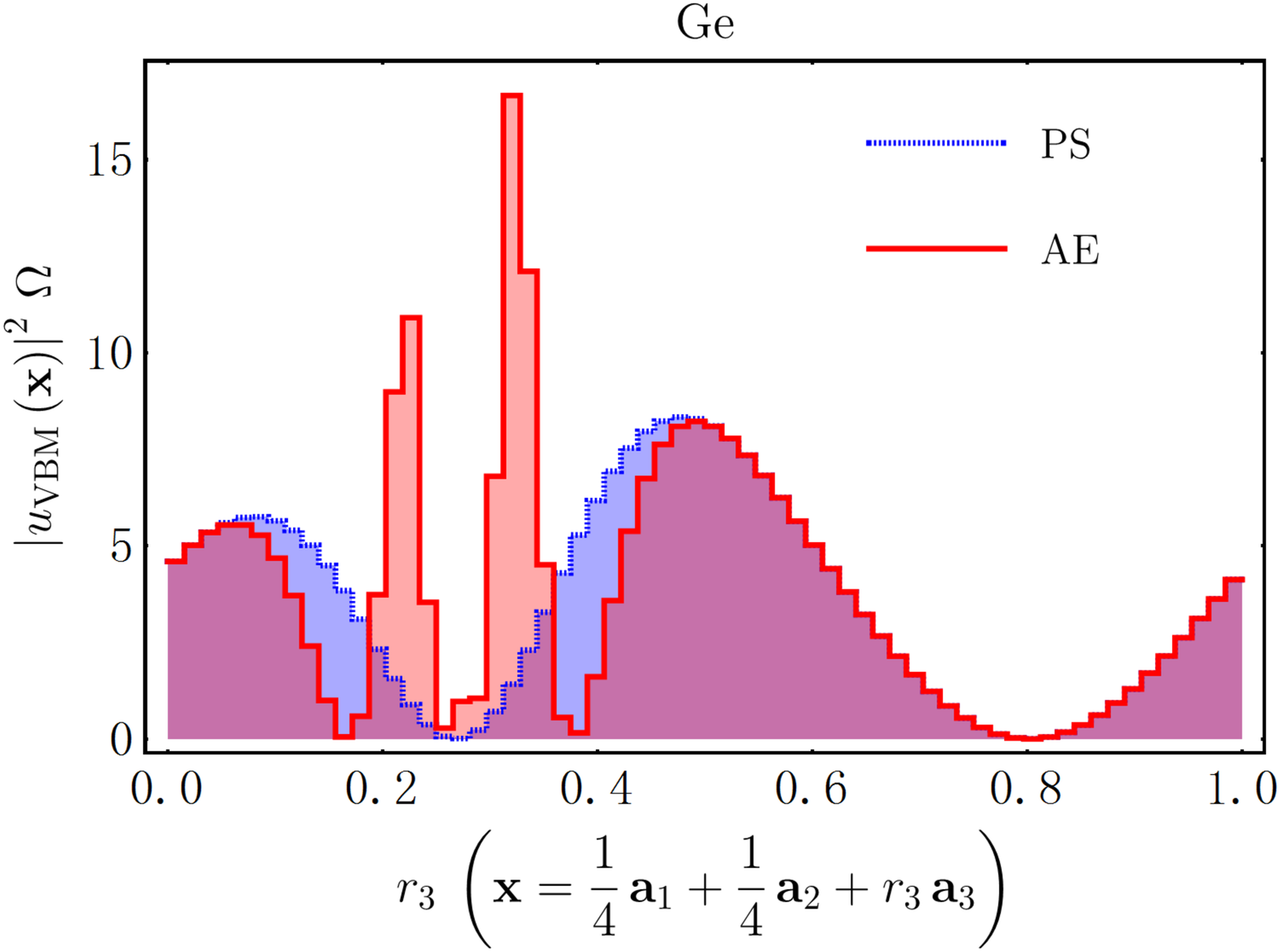}
\par\end{centering}

\begin{centering}
\includegraphics[scale=0.2]{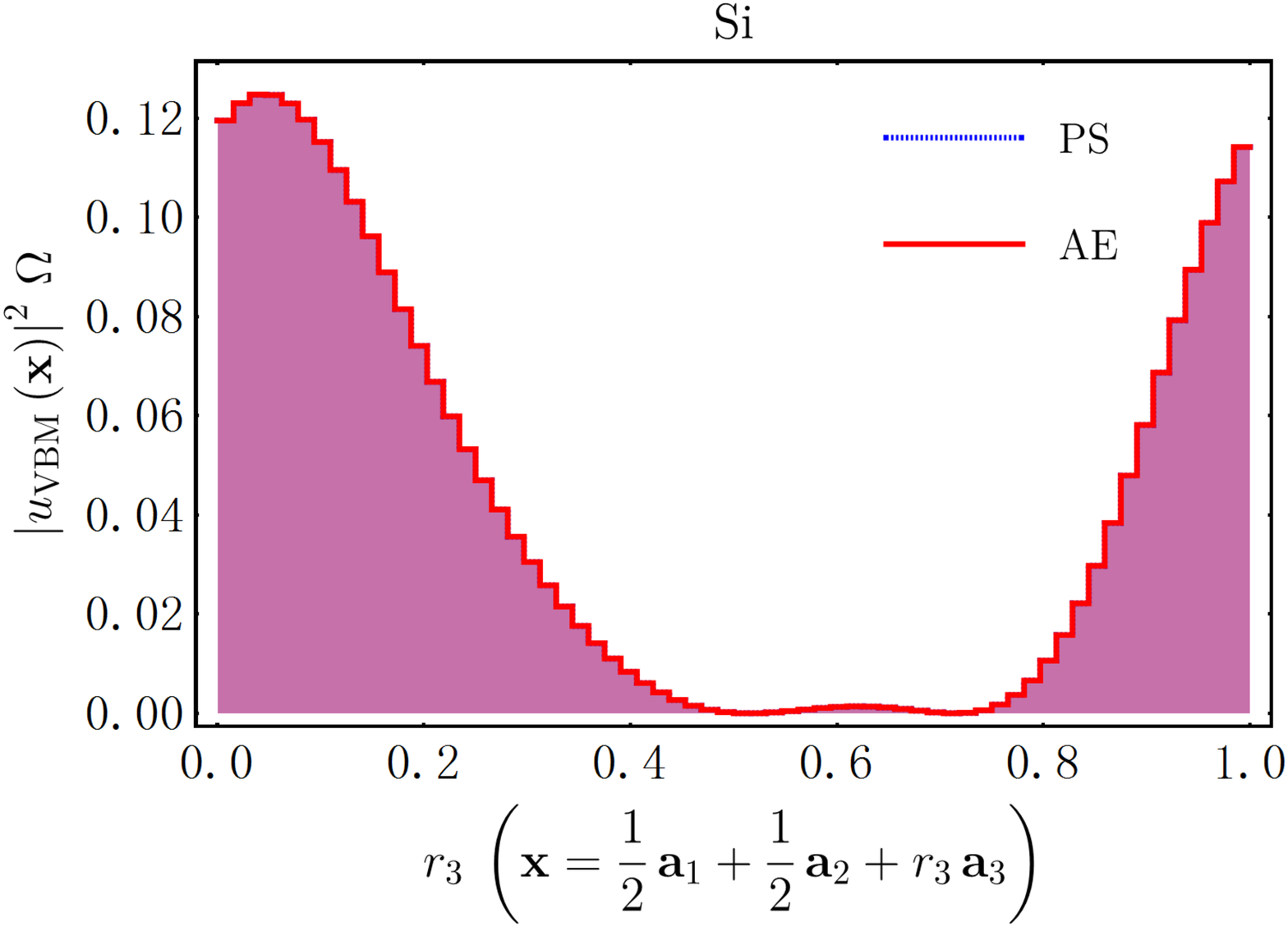}$\qquad$\includegraphics[scale=0.2]{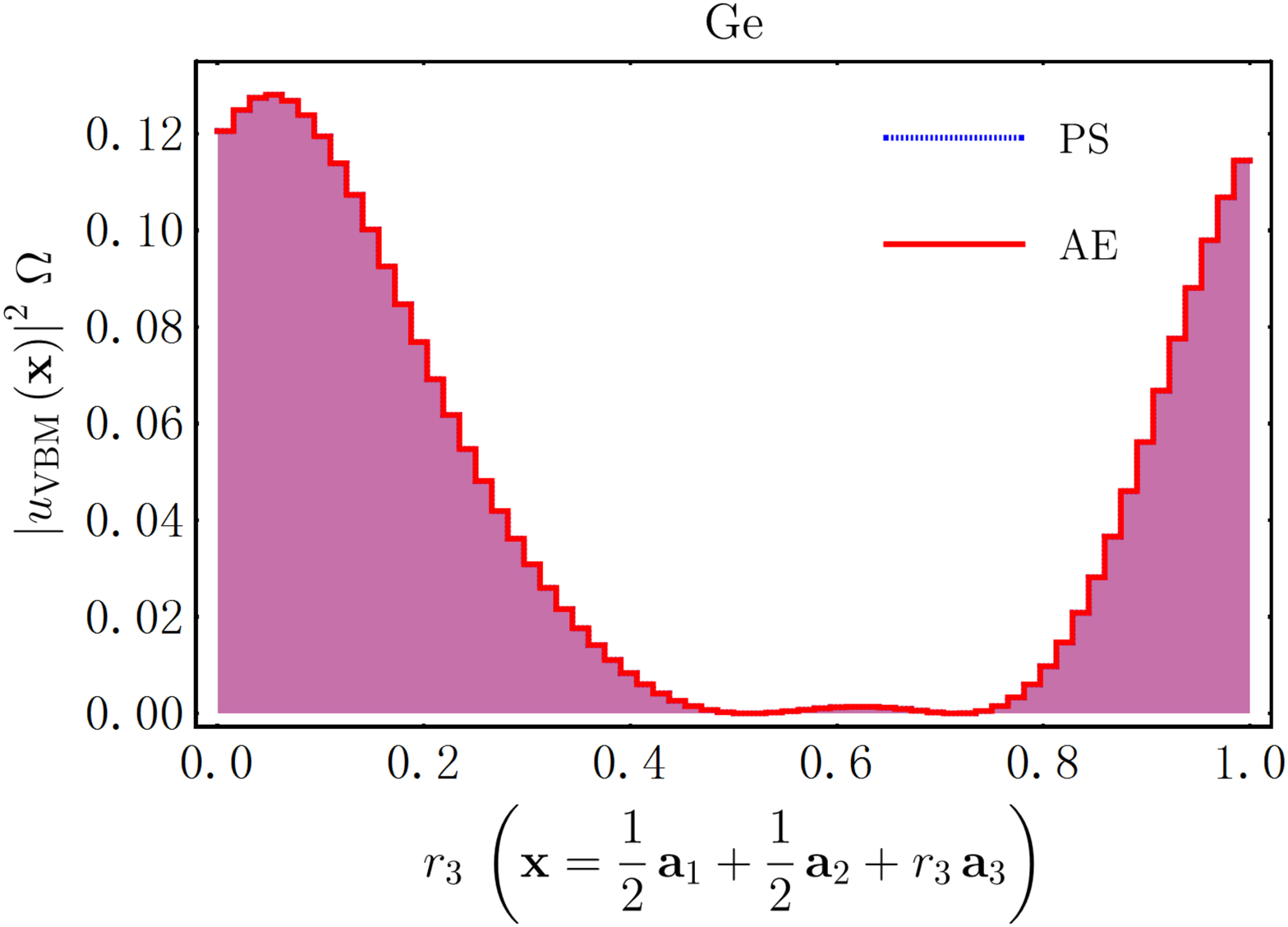}
\par\end{centering}

\protect\caption{\label{fig:Electron Density}Comparison of the valence electron densities
of the highest valence band state at $\Gamma$, for silicon (\textit{left})
and germanium (\textit{right}), determined from the AE (\textit{red}) and PS
(\textit{blue}) approaches within the PAW framework, respectively.
In the \textit{upper} row, the densities are along the path $\mathbf{x}=r_{3}\,\mathbf{\mathbf{a}}_{3}$~$\left(0\leq r_{3}<1\right)$
inside the UC, while in \textit{middle} and the \textit{bottom} rows
presented are those along the paths $\mathbf{x}=\frac{1}{4}\,\mathbf{a}_{1}+\frac{1}{4}\,\mathbf{a}_{2}+r_{3}\,\mathbf{a}_{3}$
and $\mathbf{x}=\frac{1}{2}\,\mathbf{a}_{1}+\frac{1}{2}\,\mathbf{a}_{2}+r_{3}\,\mathbf{a}_{3}$
, respectively. See text for details.}
\end{figure}

Fifteen adjacent bands from both the valence and conduction states
across the band gaps are presented in Fig.~\ref{fig:band structure}
for silicon and germanium, respectively, along a specific Brillouin
zone path $L$-$\Gamma$-$X$-$\Gamma$. The \textit{valence band
maximums} (VBM) are rescaled to $0\mathrm{\, eV}$ for convenience,
and the band-gap energies $E_{\mathrm{g}}$, namely, the energies of
the \textit{conduction band minimums} (CBM) are calibrated to the
experimental values for silicon ($1.12\,\mathrm{eV}$) and germanium
($0.67\,\mathrm{eV}$), respectively, using the scissor operation.
In calculation, we adopt an energy cutoff $E_{\mathrm{cut}}=96\,\mathrm{Ry}$,
a value sufficiently large to obtain  converged band structures in
the PAW framework. This energy cutoff translates to a cap on the radius
of the reciprocal vector $\mathbf{G}$ at $G_{\mathrm{max}}=\sqrt{2m_{e}E_{\mathrm{cut}}}$.
Moreover, the \textit{fast Fourier transformation} (FFT) of the electronic
density, or square of the wave functions, which makes its appearance
in Eq.~(\ref{eq:total ExcitationRate}), requires PW components up
to a reciprocal vector radius twice as large as $G_{\mathrm{max}}$ in order to avoid the wrap-around errors. So in practice the charge
density is presented in a box consisting of $\left(2G_{\mathrm{max}}a/\pi\right)^{3}\approx64^{3}$
grids. In this work we use the experimental lattice constants  for silicon $a_{\mathrm{Si}}=10.263\,\mathrm{Bohr}$
and germanium $a_{\mathrm{Ge}}=10.619\,\mathrm{Bohr}$, respectively.

In addition, to illustrate the source of possible error from the pseudopotential
method for the calculation of transition event rate Eq.~(\ref{eq:total ExcitationRate}),
in Fig.~\ref{fig:Electron Density} we make an elaborate comparison
of the valence electron densities between the AE and PS approaches
inside the UC. For concreteness, we take the VBM (at the $\Gamma$ point $\mathbf{k}=\mathbf{0}$) for instance,
and choose three specific paths within the UC of the two crystals,
which are $\mathbf{x}=r_{3}\,\mathbf{\mathbf{a}}_{3}$ (in the upper
row), $\mathbf{x}=\frac{1}{4}\,\mathbf{a}_{1}+\frac{1}{4}\,\mathbf{a}_{2}+r_{3}\,\mathbf{a}_{3}$
(in the middle row), and $\mathbf{x}=\frac{1}{2}\,\mathbf{a}_{1}+\frac{1}{2}\,\mathbf{a}_{2}+r_{3}\,\mathbf{a}_{3}$
(in the bottom row) with $0\leq r_{3}<1$, respectively,
where the primitive vectors $\left\{ \mathbf{\mathbf{a}}_{1},\mathbf{\mathbf{a}}_{2},\mathbf{\mathbf{a}}_{3}\right\} =\left\{ a\left(0,\frac{1}{2},\frac{1}{2}\right),\, a\left(\frac{1}{2},0,\frac{1}{2}\right),\, a\left(\frac{1}{2},\frac{1}{2},0\right)\right\} $
are expressed in Cartesian coordinates and in terms of the lattice
constant $a$. Corresponding electronic densities are normalized within
the UC (see Eq.~(\ref{eq:PeriodWaveFunction})) with condition $\ensuremath{\int\left|u_{\mathrm{VBM}}\left(\mathbf{x}\right)\right|^{2}\Omega\,\mathrm{d}^{3}r=1}
 $.  As shown in the first and second rows in Fig.~\ref{fig:Electron Density},
the AE electronic density (red) features rapid oscillations when the
path passes through the nuclei, while outside the augmentation radius
the PS density (blue) well coincides with the AE one. When the path
does not intersect the augmentation sphere, as shown in the third
row, the two wave functions are completely equal.

Another observation from Fig.~\ref{fig:Electron Density} is that
the oscillatory behavior of the wave function near the nucleus is
so remarkable that the present real grids fall short of resolution
to continuously depict the variation details. If the kinetics allows
for a large transferred momentum in transition, these highly oscillating
wave functions may couple with the high-frequency phase factor terms
\{$e^{-i\mathbf{\mathbf{G}}\cdot\mathbf{x}}$\} in the expression
of transition rate Eq.~(\ref{eq:total ExcitationRate}), and hence
more $\mathbf{G}$-vector terms can contribute to the total transition
rate. The maximum transferred momentum for the given energy difference
$E_{i'\mathbf{k}'}-E_{i\mathbf{k}}$ and DM incident velocity $w$
can be drawn from the Heaviside function $\varTheta$ in Eq.~(\ref{eq:total ExcitationRate})
as the following,
\begin{eqnarray}
q_{\mathrm{max}}\left(E_{i'\mathbf{k}'}-E_{i\mathbf{k}}\right) & = & m_{\chi}w\left[1+\sqrt{1-\frac{2\left(E_{i'\mathbf{k}'}-E_{i\mathbf{k}}\right)}{m_{\chi}\, w^{2}}}\right].\label{eq:q-relation}
\end{eqnarray}
In contrast to the case of optical excitation, this massive particle
kinetic relation translates to an average momentum transfer $q$ of
$\mathcal{O}\left(c/w_{0}\right)\approx\mathcal{O}\left(10^{3}\right)$
larger given the same transition energy, where $w_{0}=300\,\mathrm{km\cdot s}^{-1}$
represents the typical speed of the galactic DM particles. As a consequence,
a larger modulus of $\mathbf{G}$-vectors $G$ is accordingly required
to resolve the short-range details in this deep inelastic scattering
process. To get a sense, we take the excitation channel VBM$\rightarrow$CBM
in silicon crystal for instance, where a $50\,\mathrm{MeV}$-DM particle
with a typical velocity $w_{0}=300\,\mathrm{km\cdot s}^{-1}$ can
produce a transferred momentum as large as $0.099\,\mathrm{MeV}$,
which corresponds to an energy cutoff $E_{\mathrm{cut}}\approx176\,\mathrm{Ry}$,
far exceeding the typical value of $30\sim60\,\mathrm{Ry}$ in pseudopotential
PW calculations. While these large-$G$ phase factors \{$e^{-i\mathbf{\mathbf{G}}\cdot\mathbf{x}}$\}
decouple from the smooth PS wave functions, they may still resonate
with the rapidly oscillatory AE ones.

\begin{figure}
\begin{centering}
\includegraphics[scale=0.4]{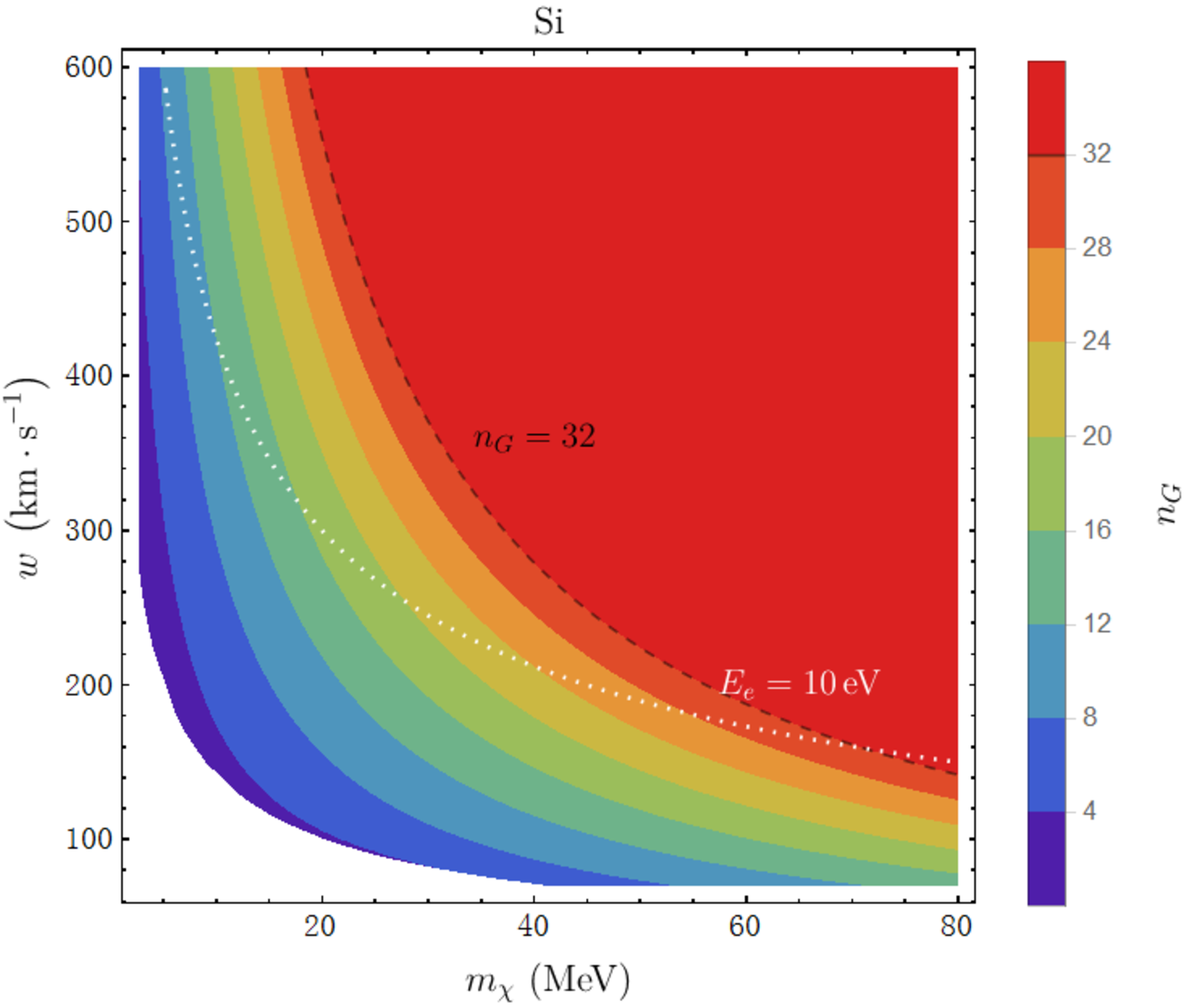}$\quad$\includegraphics[scale=0.4]{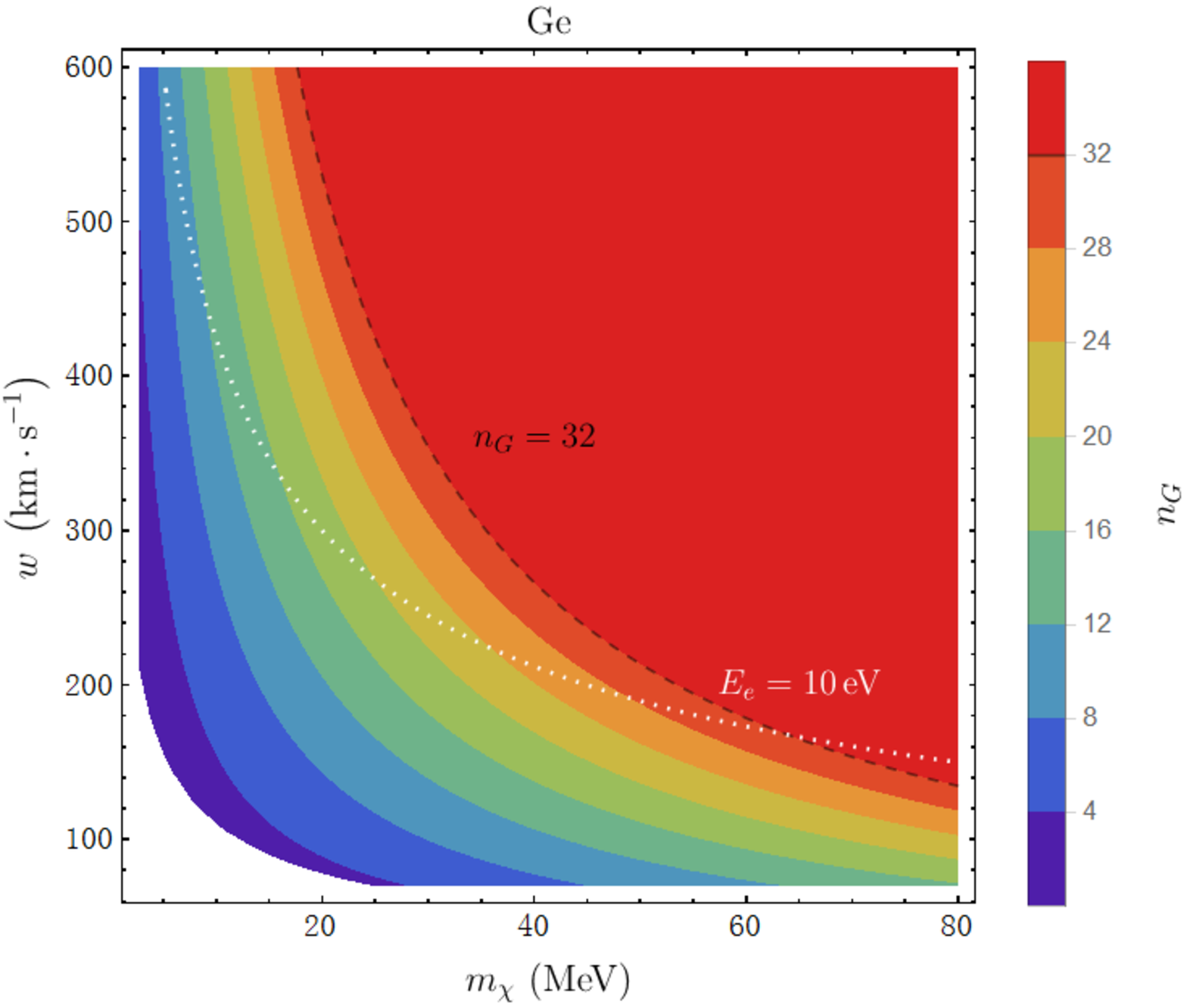}\vspace{0.25cm}

\par\end{centering}

\begin{centering}
\includegraphics[scale=0.4]{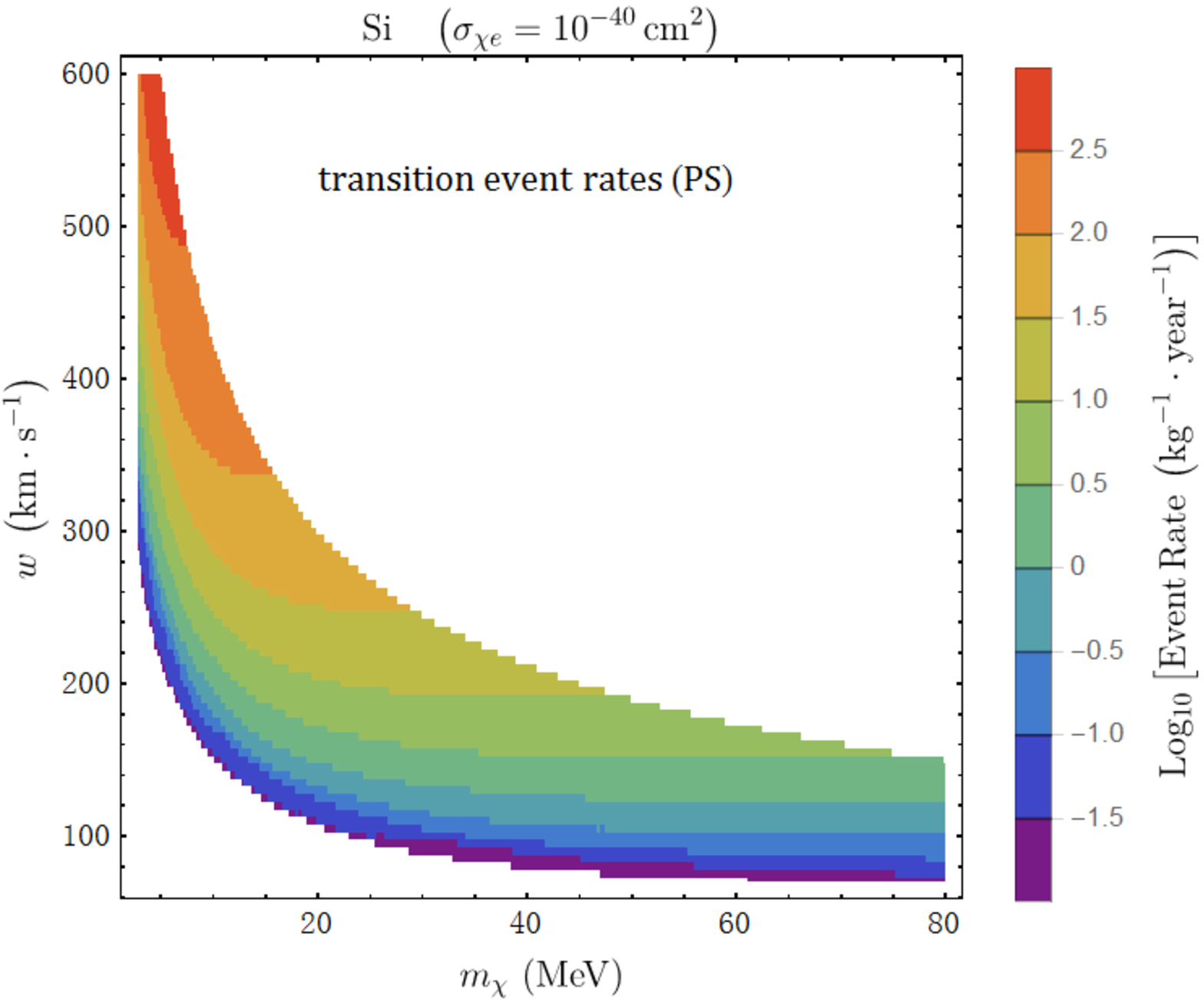}$\quad$\includegraphics[scale=0.4]{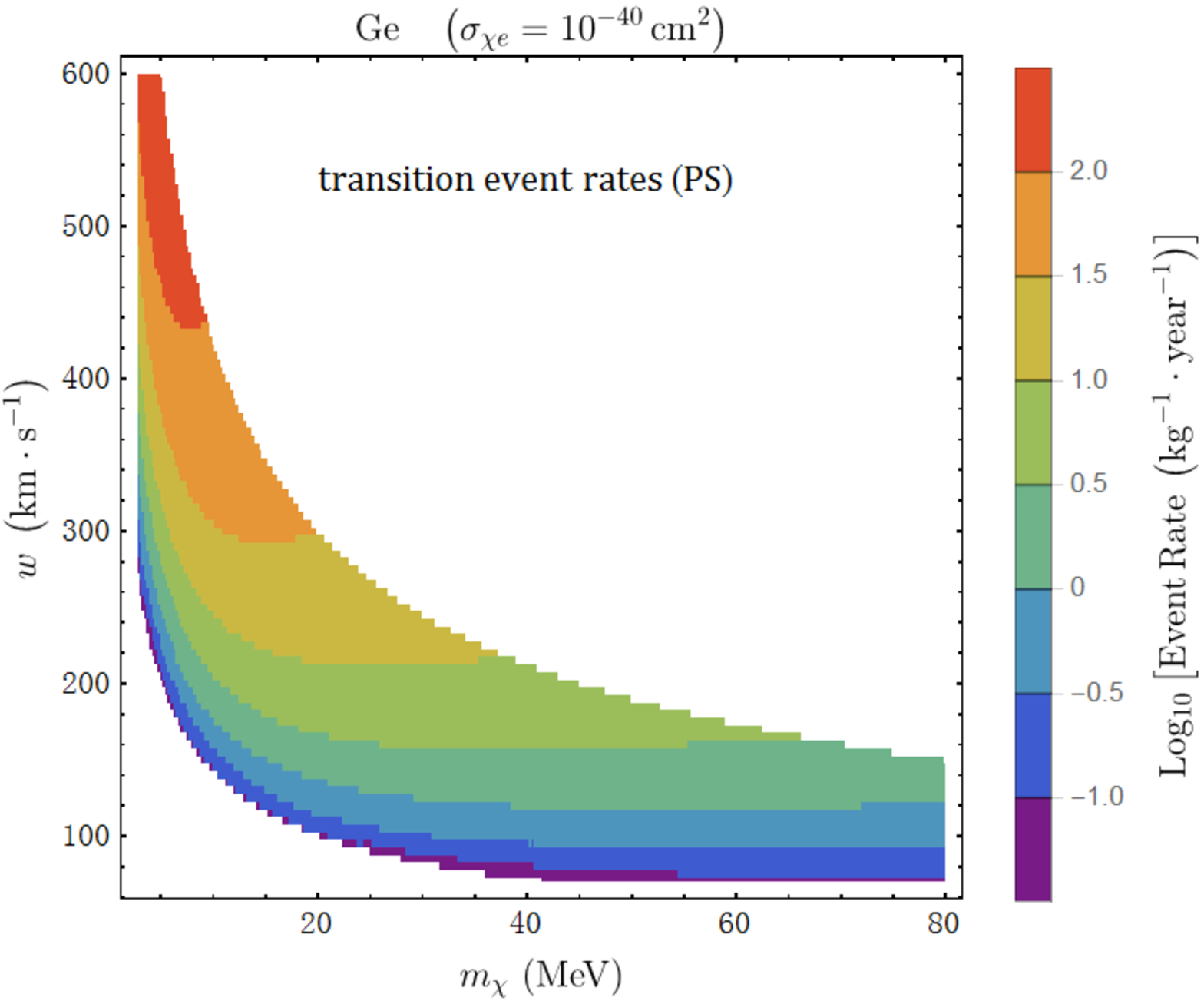}\vspace{0.25cm}

\par\end{centering}

\begin{centering}
\includegraphics[scale=0.4]{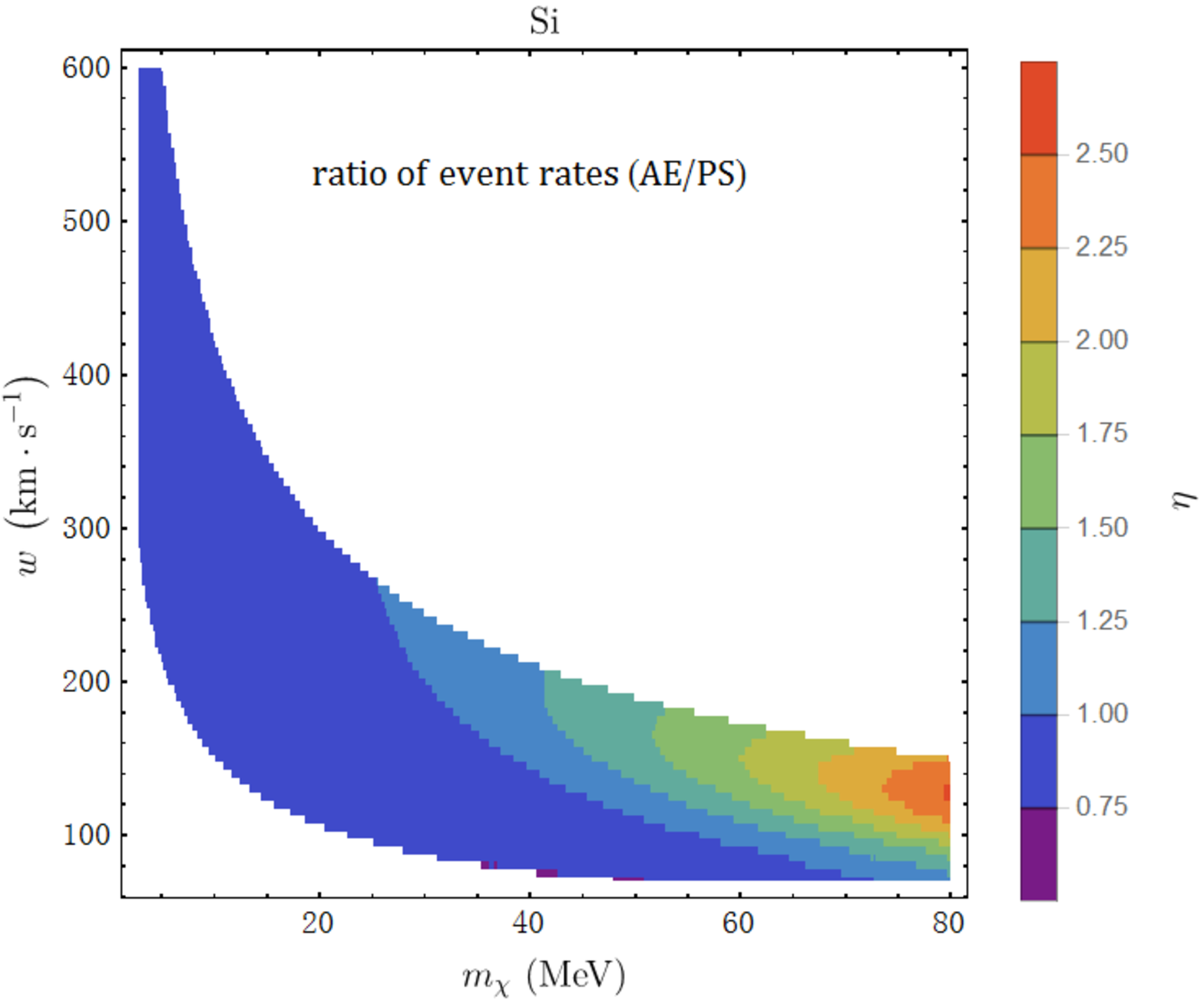}$\quad$\includegraphics[scale=0.4]{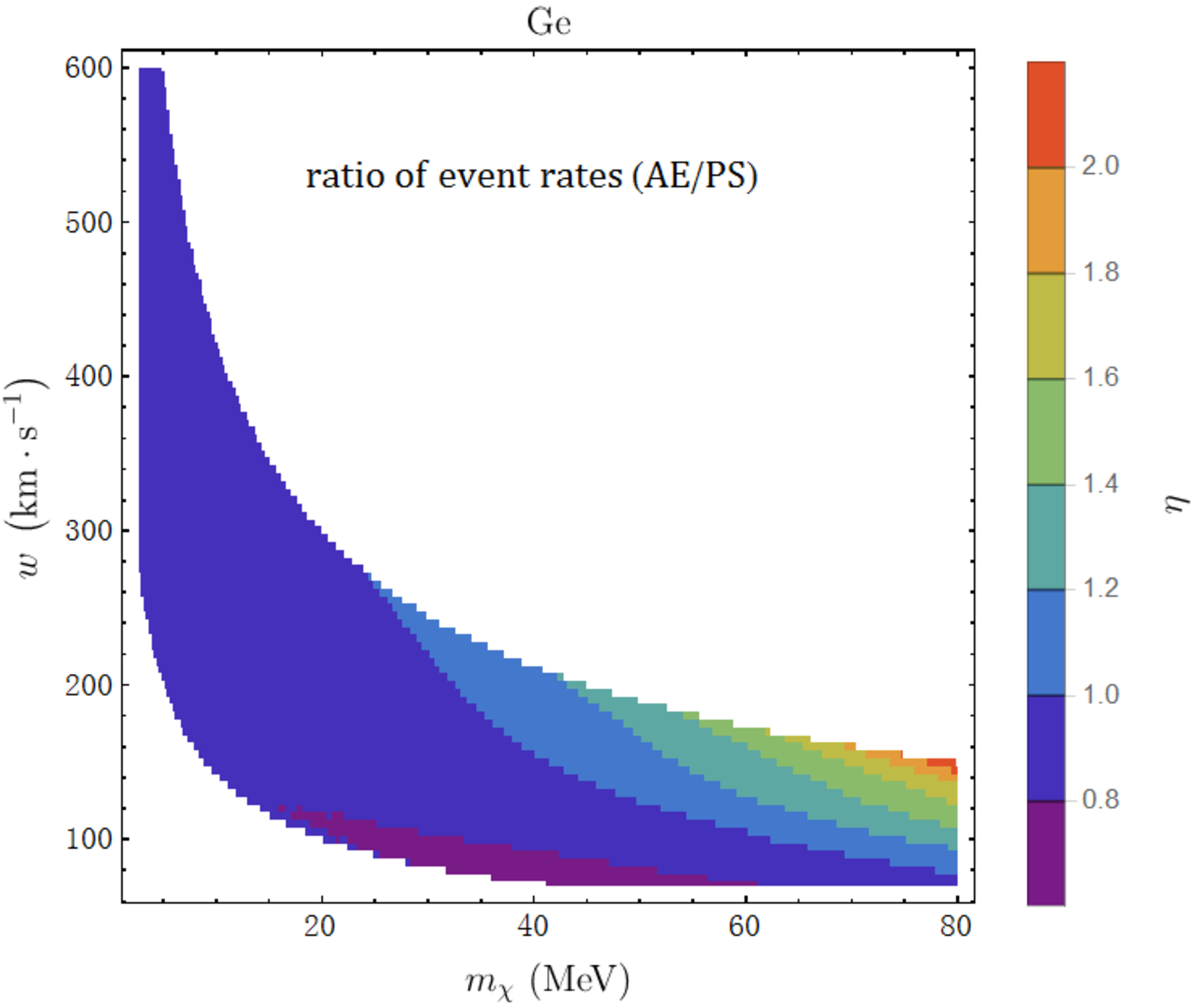}
\par\end{centering}

\protect\caption{\label{fig:Gcut} \textbf{\textit{Upper}}: The maximum modulus number
of the reciprocal $\mathbf{G}$-vectors involved in the DM-induced
transition events for parameters $(m_{\chi},\, w)$, for silicon (\textit{left}),
and germanium (\textit{right}). The black dashed lines mark the truncation
$n_{G}\leq32$ imposed by the grid resolution in the real space, while
the white dotted lines represent the transition energy bound $E_{e}=10\,\mathrm{eV}$
adopted in this work. \textbf{\textit{Middle}}: The calculated DM-induced
transition events for silicon (\textit{left}), and germanium (\textit{right}),
for a DM-electron cross section $\sigma_{\chi e}=10^{-40}\,\mathrm{cm}^{2}$.\textbf{\textit{
Bottom}}: Ratios of the AE event rate to the PS event rate for silicon
(\textit{left}) and germanium (\textit{right}), respectively. See
text for details.}

\end{figure}

Therefore, in order to investigate the AE reconstruction effects on
the calculation of DM-induced transition rate, we need to choose an
appropriate set of parameters in a sense
that the maximum momentum transfer should be compatible with the resolution
of the real grid adopted in computation. For illustration, we present the
largest modulus of  integer number vectors
for the diamond structure, \textit{i.e.}, $n_{G}=q_{\mathrm{max}}\left(E_{\mathrm{g}}\right)\, a/(2\pi)$~\footnote[1]{\renewcommand{\baselinestretch}{1}\selectfont Given the constraint $\left|n_{1}\mathbf{b}_{1}+n_{2}\mathbf{b}_{2}+n_{3}\mathbf{b}_{3}\right|\leq q_{\mathrm{max}}\left(E_{\mathrm{g}}\right)$, where  $\left\{ \mathbf{b}_{i}\right\}$ are the primitive reciprocal vectors  and $\left\{ n_{i}\right\}$  are a set of integer numbers, the largest modulus $n_{G}\equiv\left|\mathbf{n}\right|
 =\left(n_{1}^{2}+n_{2}^{2}+n_{3}^{2}\right)^{1/2}$ simply equals $q_{\mathrm{max}}\left(E_{\mathrm{g}}\right)\, a/(2\pi)$ for the diamond structure.} versus
two parameters: DM mass $m_{\chi}$ and incident velocity $w$, for
silicon and germanium crystals in the upper row in Fig.~\ref{fig:Gcut}.
The kinetic energy cutoff $E_{\mathrm{cut}}=96\,\mathrm{Ry}$
adopted in this study  corresponds to the constraint
$n_{G}\leq32$ (in black dashed line) in the $\mathtt{ABINIT}$ implementation of PAW. That
is to say, the resolution associated with the  cutoff $E_{\mathrm{cut}}=96\,\mathrm{Ry}$
can only describe  the electronic excitation processes in
colored parameter regions below the black dashed lines in Fig.~\ref{fig:Gcut},
while in the white areas corresponding excitations are kinetically
forbidden.

After determining the parameter region compatible with the given energy
cutoff, now we can compute the DM-induced electronic excitation rates
for the silicon and germanium semiconductor detectors. In calculation
we truncate the energy transfer $E_{e}$ at $10\,\mathrm{eV}$, which
corresponds to including only the transition events from the 4 valence
bands to the 7 lowest conduction bands. Such constraint is presented in
white dotted lines in the upper row in Fig.~\ref{fig:Gcut} for silicon
and germanium, respectively. Parameters $(m_{\chi},\, w)$ above the
white line can induce excitation events with transition energies above $10\,\mathrm{eV}$, and hence more unoccupied energy bands are
required for exact calculation. In addition, although  the transition rate Eq.~(\ref{eq:total ExcitationRate})
depends on the orientation of the crystal target with respect to galaxy,
for simplicity however, in calculation the velocity distribution is
approximated as an isotropic one for parameter $(m_{\chi},\, w)$,\textit{
i.e.}, $g_{\chi}\left(\mathbf{v},\,\hat{\mathbf{q}}\right)=\left(4\pi\right)^{-1}\delta\left(v-w\right)$.
Besides, following the strategy in Ref.~\cite{Essig:2015cda}, we
equivalently express Eq.~(\ref{eq:total ExcitationRate}) in terms
of transferred momentum $q$ and energy transfer $E_{e}$ in order
to make the scan of the parameter space computationally more efficient.
In short, after inserting the monochromatic velocity distribution
the total transition rate for $(m_{\chi},\, w)$ can be recast as
\begin{eqnarray}
\mathcal{R}\left(m_{\chi},\, w\right) & = & \frac{\rho_{\chi}}{m_{\chi}}\left(\frac{2\pi^{2}\sigma_{\chi e}}{w\,\mu_{\chi e}^{2}\, q_{\mathrm{ref}}}\right)\frac{N_{\mathrm{cell}}}{\varOmega}\int\frac{\mathrm{d}E_{e}}{E_{\mathrm{ref}}}\,\frac{\mathrm{d}q}{q_{\mathrm{ref}}}\,\Theta\left[w-w_{\mathrm{min}}\left(q,\, E_{e}\right)\right]\,\mathcal{F}\left(q,\, E_{e}\right),\label{eq:EvenRateScan}
\end{eqnarray}
for either the AE  or the PS wavefunctions, where  two reference values for momentum and energy are constructed
as $q_{\mathrm{ref}}=2\pi/a$ and $E_{\mathrm{ref}}=q_{\mathrm{ref}}^{2}/\left(2\, m_{e}\right)$,
respectively, $N_{\mathrm{cell}}=V/\varOmega$ is the number of cells
in the crystal target, and the non-dimensional form factor is introduced
as
\begin{eqnarray}
\mathcal{F}\left(q,\, E_{e}\right) & \equiv & \left(\frac{q_{\mathrm{ref}}}{q}\right)\sum_{i'}^{c}\sum_{i}^{v}\int_{\mathrm{BZ}}\frac{\varOmega\,\mathrm{d}^{3}k'}{\left(2\pi\right)^{3}}\frac{\varOmega\,\mathrm{d}^{3}k}{\left(2\pi\right)^{3}}\,\Bigg\{ E_{\mathrm{ref}}\,\delta\left(E_{i'\mathbf{k}'}-E_{i\mathbf{k}}-E_{e}\right)\nonumber \\
 &  & \times\left.\sum_{\mathbf{G}}\left|\int_{\Omega}\mathrm{d}^{3}x\, u_{i'\mathbf{k}'}^{*}\left(\mathbf{x}\right)\, e^{-i\mathbf{\mathbf{G}}\cdot\mathbf{x}}\, u_{i\mathbf{k}}\left(\mathbf{x}\right)\right|^{2}q_{\mathrm{ref}}\,\delta\left(\left|\mathbf{k}-\mathbf{k}'+\mathbf{G}\right|-q\right)\right\} \label{eq:FormFactor}.
\end{eqnarray}
 In realistic computation, several numerical modifications are made
to Eq.~(\ref{eq:EvenRateScan}) as follows.
\begin{itemize}
\item The integrals over $q$ and $E_{e}$ are replaced by the summations
over uniform discrete bins of $q$ and $E_{e}$. We use the bin widths
$\Delta E=0.058\,\mathrm{eV}$ and $\Delta q=78.6\,\mathrm{eV}$,
and the integrand is evaluated at the central value in each relevant
bin. The delta functions are discretized to a normalized step function.
For example, if a specific parameter $E_{e}$ falls into the $n$-th
energy bin, then the function $\delta\left(E_{i'\mathbf{k}'}-E_{i\mathbf{k}}-E_{e}\right)$
is approximated as $\Delta E^{-1}\cdot\Theta\left(\frac{\Delta E}{2}-\left|E_{i'\mathbf{k}'}-E_{i\mathbf{k}}-E_{n}\right|\right)$,
with $E_{n}$ being the central value of this bin.
\item As a standard DFT numerical procedure, the integrals of the continuous
$k$-points in the first BZ are replaced by the summations over a
uniform discrete mesh of representative $k$-points in the following
manner:
\begin{eqnarray}
\int_{\mathrm{BZ}}\frac{\varOmega\,\mathrm{d}^{3}k}{\left(2\pi\right)^{3}}\left(\cdots\right) & \rightarrow & \frac{1}{N_{k}}\sum_{\mathbf{k}}^{N_{k}}\left(\cdots\right),
\end{eqnarray}
where $N_{k}$ is the number of $k$-points sampled in the first BZ.
In this study, we use a homogeneous set of $6\times6\times6$ $k$-points.
\item Integral $\int_{\Omega}\mathrm{d}^{3}x\, u_{i'\mathbf{k}'}^{*}\left(\mathbf{x}\right)\, e^{-i\mathbf{\mathbf{G}}\cdot\mathbf{x}}\, u_{i\mathbf{k}}\left(\mathbf{x}\right)$
can be understood as the Fourier transformation of the square term
$u_{i'\mathbf{k}'}^{*}\left(\mathbf{x}\right)\, u_{i\mathbf{k}}\left(\mathbf{x}\right)$
within the UC, which is practically performed using the discrete FFT
technique, so the reciprocal $\mathbf{G}$-vectors have a natural
truncation consistent with the grid number in the real space. As mentioned
above, the modulus number of the $\mathbf{G}$-vectors $n_{G}$ can
extend to a value no more than 32 in our analysis.
\end{itemize}
In the middle row of Fig.~\ref{fig:Gcut}, we present the  calculated
transition event rates per kilogram-day from the PS wavefunctions
for silicon (left) and germanium (right), respectively.
A DM-electron cross section $\sigma_{\chi e}=10^{-40}\,\mathrm{cm}^{2}$
is assumed, and only the parameter regions below the transition energy
constraint (white dotted line in the upper row) and the $\mathbf{G}$-vector
bound (black dashed line in the upper row) are calculated. In the
bottom row shown are the comparisons between the theoretical excitation
event rates derived from the AE and PS approaches for the two crystals,
in terms of the ratio of the AE event rate to its PS counterpart for
each parameter pair $m_{\chi}$ and $w$, \textit{i.e.}, $\eta\equiv\mathcal{R}_{\mathrm{AE}}\left(m_{\chi},\, w\right)/\mathcal{R}_{\mathrm{PS}}\left(m_{\chi},\, w\right)$.
In large part of the scanned region, the two calculated even rates
are found to be quite consistent. Interestingly, however, in  areas
where large momentum transfer (or equivalently, large $n_{G}$) can
be induced, the estimates based on the AE and PS approaches begin
to differ with each other, and this trend seems to continue  into larger $\mathbf{G}$-vector
parameter region. Therefore,  the actual electronic
distribution near the nucleus is of remarkable relevance for the estimation
of the DM-induced excitation rate, especially in the area where a
large momentum transfer is favored in the excitation process.
\begin{figure}
\begin{centering}
\includegraphics[scale=0.2]{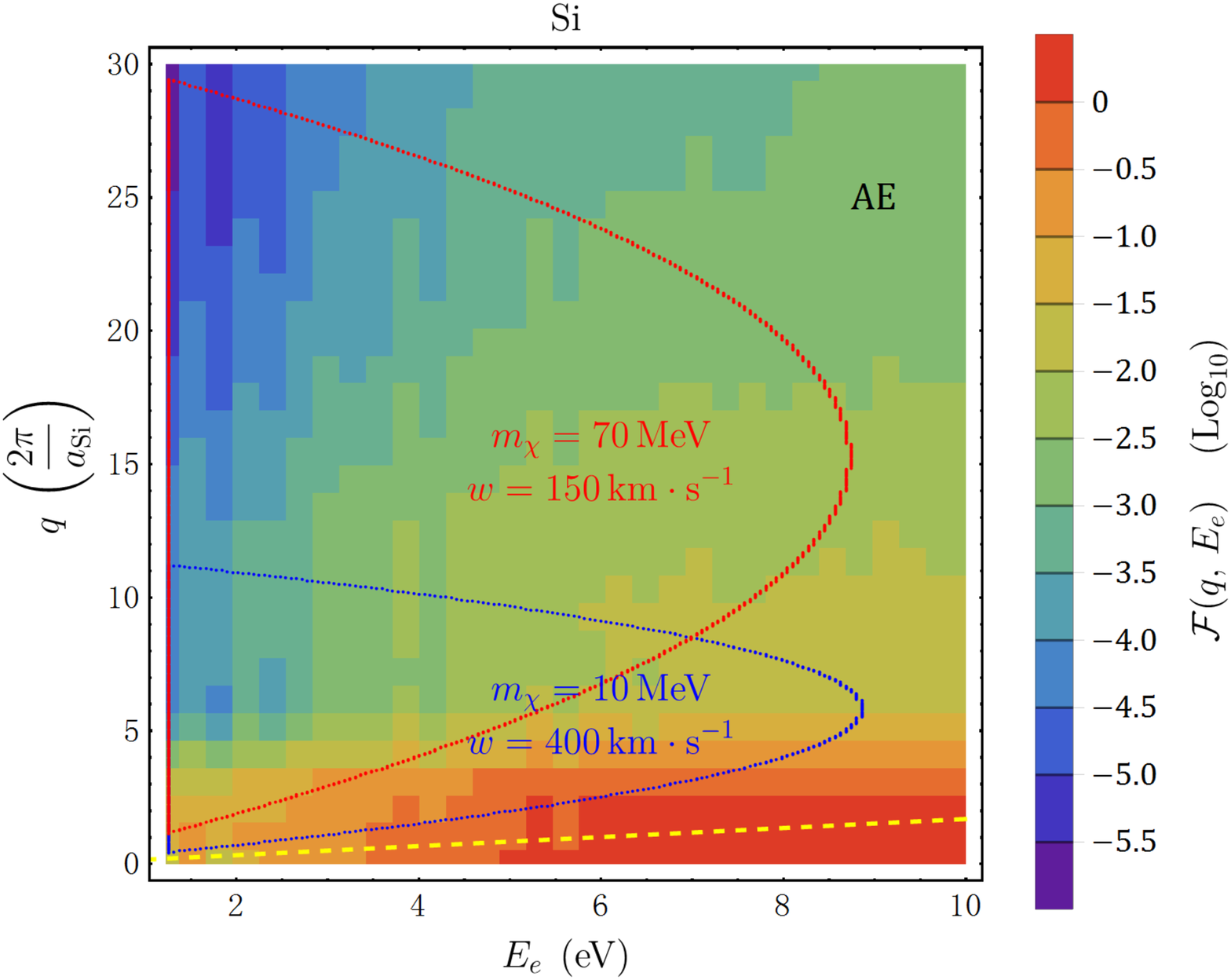}\hspace{0.2cm}\includegraphics[scale=0.2]{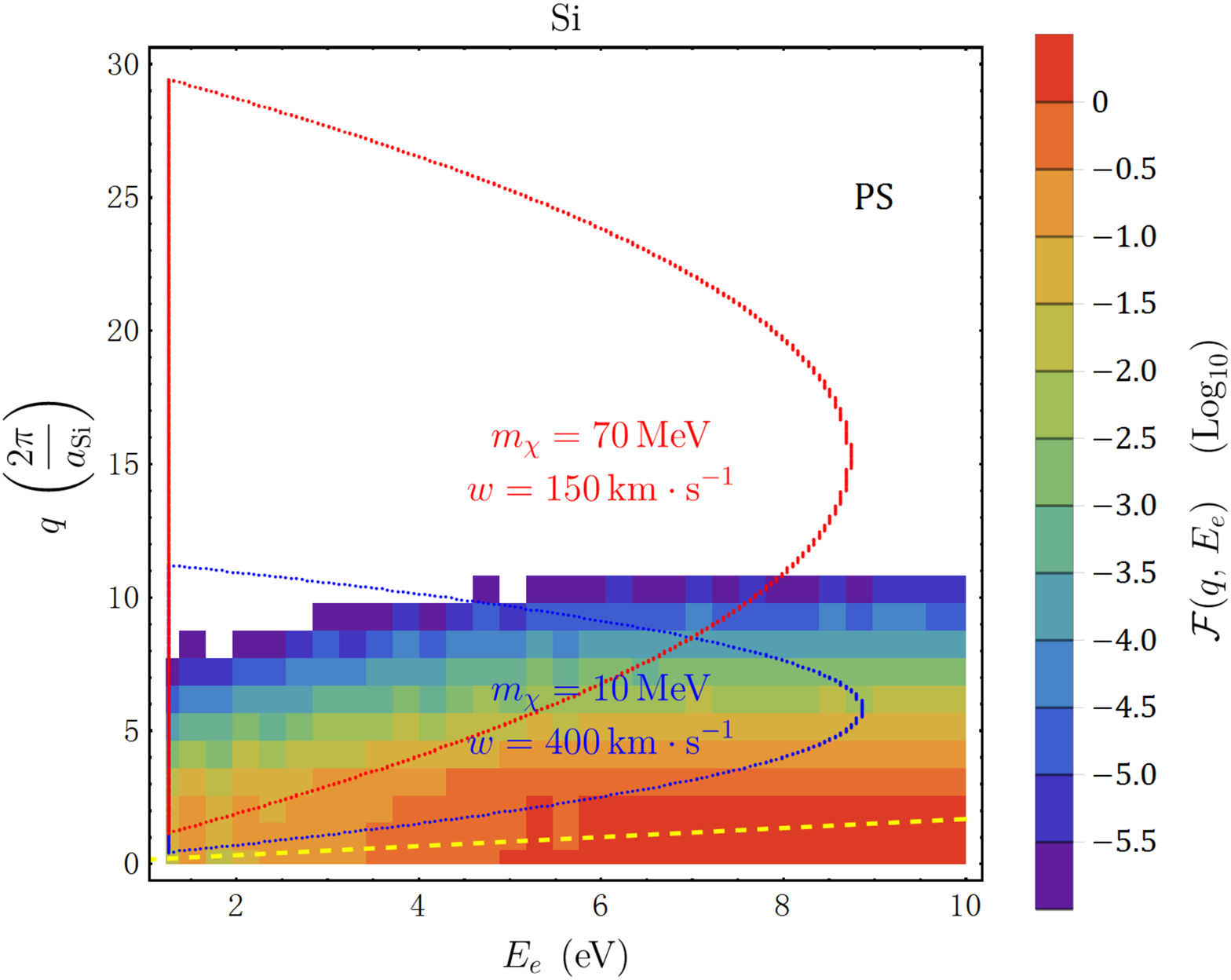}
\par\end{centering}

\protect\caption{\label{fig:qE_constract} Crystal form factor $\mathcal{F}\left(q,\, E_{e}\right)$
defined in Eq.~(\ref{eq:FormFactor}) calculated from the AE (\textit{left})
and PS (\textit{right}) approaches, respectively. The area enveloped in \textit{red}
(\textit{blue}) represents the parameter region relevant for the total
excitation event rate for parameter $m_{\chi}=70\,\mathrm{MeV}$,
$w=150\,\mathrm{km\cdot s^{-1}}$ ($m_{\chi}=10\,\mathrm{MeV}$, $w=400\,\mathrm{km\cdot s^{-1}}$). Only the region above the  yellow dashed line is kinetically allowed for the galactic DM  distribution.
See text for details.}
\end{figure}

In order to demonstrate the origin of such deviation, in Fig.~\ref{fig:qE_constract}
we present the comparison of the form factor $\mathcal{F}\left(q,\, E_{e}\right)$
in Eq.~(\ref{eq:EvenRateScan}), between the AE and PS approaches
for silicon. This non-dimensional quantity is convenient for direct
contrast because all the integrand function values in pixels are accounted
with equal weight when summed up to obtain the total event rate. The
momentum transfer $q$ is expressed in terms of the silicon reciprocal
lattice $2\pi/a_{\mathrm{Si}}$. We choose two benchmark points representative
of two distinct scenarios in the bottom row of Fig.~\ref{fig:Gcut},
\textit{i.e.}, when the AE-based estimate of event rate differs with
its PS counterpart, and when the AE-based estimate of event rate agrees
with its PS counterpart.

For the first case, we take the parameter
pair $m_{\chi}=70\,\mathrm{MeV}$, $w=150\,\mathrm{km\cdot s^{-1}}$
as example. At this point one has $\eta=2.01$, and the kinetics confines
the integration in Eq.~(\ref{eq:EvenRateScan}) within the area in
the red contour in Fig.~\ref{fig:qE_constract}, only where
the Heaviside step function yields non-zero value. In addition, an
important observation from Fig.~\ref{fig:qE_constract} is that as
we expected, the AE wavefunctions near the nucleus, reconstructed
from the PAW transformation~Eq.~(\ref{eq:ReconstructionRelation}), indeed play a non-negligible role in
transitions where large transferred momentum participates. While the
PS approach gives an accurate description of the excitation process
in the parameter area below $q\approx5\left(2\pi/a_{\mathrm{Si}}\right)$,
(which is referred to as small-$q$ region, and the rest area is referred
to as large-$q$ region, for convenience) the relevant form factor
falls off so rapidly towards the large-$q$ region that the
PS scheme may cause remarkable error in the calculation of event rates.
Comparing the AE case (left) to the PS one (right),
one finds that the large ratio $\eta$ can be attributed to the fact
that the contour on one hand covers only a small portion of the parameter
area where the PS-based description remains effective, and on the other
it contains a sizable part of the large-$q$ parameter area. These
two factors effectively enchance the contrast.

As for the other
case, we pick the parameter pair $m_{\chi}=10\,\mathrm{MeV}$, $w=400\,\mathrm{km\cdot s^{-1}}$
($\eta=0.85$) for illustration. Since half of the kinetically allowed
region (contoured in blue line) overlaps the small-$q$
area, based on the above reasoning, it is understandable that the
corrections from the AE approach are not so remarkable under this
circumstance. Moreover,  the galactic DM velocity distribution   also imposes a kinetic constraint on the $E_{e}$-${q}$ plane. Actually,  given the truncated Maxwellian distribution introduced in Sec.~\ref{sec:TransitionRate}, and the fact that the  $w_{\mathrm{min}}\left(q,\, E_{e}\right)$   reaches its minimum  as $m_{\chi}\rightarrow\infty$ (see Eq.~(\ref{eq:wmin})),  the minimal kinetically accessible transferred momentum $q$ for  recoil energy $E_{e}$ equals $E_{e}/\left(v_{\mathrm{esc}+}v_{\mathrm{e}}\right)$, which corresponds to the yellow dashed line presented in Fig.~\ref{fig:qE_constract}. The region below the line is kinetically forbidden. The implication is that large recoil energy requires large momentum transfer accordingly, and thus an accurate description of the  AE electronic wave functions becomes necessary.

\section{\label{sec:Conclusion}Conclusions and discussion}

In this work, we have investigated the implications of the PAW reconstruction
procedure on the estimation of electronic transition event rate in
two most common semiconductor targets used in the DM direct detection,
namely, silicon and germanium semiconductors. By use of the PAW method,
one can restore the AE electronic wavefunctions from the PAW PS wavefunctions
without much extra computational loads. To this end, we first utilize
the $\mathtt{atompaw}$ code to generate appropriate PAW datasets
including the PAW pseudopotential, and input these files into the
$\mathtt{ABINIT}$ package to compute the electronic structure and
relevant AE and PS wavefunctions. Based on the wavefunctions, we then
calculate relevant DM-induced excitation event rates and make comparisons
between the AE and PS approaches. It is found that the PS approach
turns inapplicable in describing excitation processes involving large
transferred momentum, because the oscillating components that would
couple with large momentum transfer are deliberately spared in calculation.
Even in the limited parameter region scanned in our study, the two
calculated event rates are found to differ with a factor of two. So
this work serves as a caveat in relevant calculations that an accurate
estimate of the DM-induced transition event rate should better be
implemented with the AE approaches.

Similar experimental strategies can be extended to the detection of the sub-MeV DM particles. For instance, in Ref.~\cite{Hochberg:2017wce} the authors discussed the theoretical prospects for detecting sub-MeV DM particles by use of the three-dimensional Dirac materials with typical band gap $E_{\mathrm{g}}\sim$
 meV. Discussion of the AE reconstruction effects can also be applied parallelly to this case. However, considering the dependence of maximum momentum transfer on the DM mass at the sub-MeV scale (see Eq.~(\ref{eq:q-relation})), the high-momentum PWs from reconstruction do not contribute to the transition event rates, so in such case the reconstruction effects are expected to be insubstantial.

Finally, a thorough estimate of the total event rate should involve
convoluting with the realistic DM velocity distribution, and thus
requires a broader recoil energy range and a larger energy cutoff
(see Fig.~\ref{fig:Gcut}), which unfortunately, is beyond the capabilities
of our cottage-industry work style. As a more efficient solution,
computation of the crystal form factor should be processed within
the well-established software packages, \textit{e.g.,} $\mathtt{ABINIT}$
in this study. We leave it for future work.

\appendix

\begin{acknowledgments}
 ZLL thanks Xingyu Gao for useful discussions. PZ is supported by the Science Challenge Project No.~TZ2016001, and NSFC No.~11625415. FZ acknowledges the support by China PFCAEP under the Grant No.~YZJJLX2016010. The computations are supported by Beijing Computational Science Research Center~(CSRC) and HPC Cluster of ITP, CAS.
\end{acknowledgments}

\bibliographystyle{JHEP}
\addcontentsline{toc}{section}{\refname}\bibliography{DM_Semiconductor}

\providecommand{\href}[2]{#2}\begingroup\raggedright\begin{thebibliography}{10}

\bibitem{Goodman:1984dc}
M.~W. Goodman and E.~Witten, {\it {Detectability of Certain Dark Matter
  Candidates}},  {\em Phys. Rev.} {\bf D31} (1985) 3059. [,325(1984)].

\bibitem{Aprile:2016swn}
{\bf XENON100} , E.~Aprile et~al., {\it {XENON100 Dark Matter Results from a
  Combination of 477 Live Days}},  {\em Phys. Rev.} {\bf D94} (2016), no.~12
  122001, [\href{http://arxiv.org/abs/1609.06154}{ arXiv:1609.06154}].

\bibitem{Akerib:2016vxi}
{\bf LUX} , D.~S. Akerib et~al., {\it {Results from a search for dark matter in
  the complete LUX exposure}},  {\em Phys. Rev. Lett.} {\bf 118} (2017), no.~2
  021303, [\href{http://arxiv.org/abs/1608.07648}{ arXiv:1608.07648}].

\bibitem{Tan:2016zwf}
{\bf PandaX-II} , A.~Tan et~al., {\it {Dark Matter Results from First 98.7 Days
  of Data from the PandaX-II Experiment}},  {\em Phys. Rev. Lett.} {\bf 117}
  (2016), no.~12 121303, [\href{http://arxiv.org/abs/1607.07400}{
  arXiv:1607.07400}].

\bibitem{Arnaud:2017usi}
{\bf EDELWEISS} , Q.~Arnaud et~al., {\it {Optimizing EDELWEISS detectors for
  low-mass WIMP searches}},  {\em Phys. Rev.} {\bf D97} (2018), no.~2 022003,
  [\href{http://arxiv.org/abs/1707.04308}{ arXiv:1707.04308}].

\bibitem{Agnese:2014aze}
{\bf SuperCDMS} , R.~Agnese et~al., {\it {Search for Low-Mass Weakly
  Interacting Massive Particles with SuperCDMS}},  {\em Phys. Rev. Lett.} {\bf
  112} (2014), no.~24 241302, [\href{http://arxiv.org/abs/1402.7137}{
  arXiv:1402.7137}].

\bibitem{Essig:2011nj}
R.~Essig, J.~Mardon, and T.~Volansky, {\it {Direct Detection of Sub-GeV Dark
  Matter}},  {\em Phys. Rev.} {\bf D85} (2012) 076007,
  [\href{http://arxiv.org/abs/1108.5383}{ arXiv:1108.5383}].

\bibitem{Graham:2012su}
P.~W. Graham, D.~E. Kaplan, S.~Rajendran, and M.~T. Walters, {\it
  {Semiconductor Probes of Light Dark Matter}},  {\em Phys. Dark Univ.} {\bf 1}
  (2012) 32--49, [\href{http://arxiv.org/abs/1203.2531}{ arXiv:1203.2531}].

\bibitem{Essig:2015cda}
R.~Essig, M.~Fernandez-Serra, J.~Mardon, A.~Soto, T.~Volansky, and T.-T. Yu,
  {\it {Direct Detection of sub-GeV Dark Matter with Semiconductor Targets}},
  {\em JHEP} {\bf 05} (2016) 046, [\href{http://arxiv.org/abs/1509.01598}{
  arXiv:1509.01598}].

\bibitem{Lee:2015qva}
S.~K. Lee, M.~Lisanti, S.~Mishra-Sharma, and B.~R. Safdi, {\it {Modulation
  Effects in Dark Matter-Electron Scattering Experiments}},  {\em Phys. Rev.}
  {\bf D92} (2015), no.~8 083517, [\href{http://arxiv.org/abs/1508.07361}{
  arXiv:1508.07361}].

\bibitem{PhysRev.136.B864}
P.~Hohenberg and W.~Kohn, {\it Inhomogeneous electron gas},  {\em Phys. Rev.}
  {\bf 136} (Nov, 1964) B864--B871.

\bibitem{PhysRev.140.A1133}
W.~Kohn and L.~J. Sham, {\it Self-consistent equations including exchange and
  correlation effects},  {\em Phys. Rev.} {\bf 140} (Nov, 1965) A1133--A1138.

\bibitem{0953-8984-21-39-395502}
P.~Giannozzi, S.~Baroni, N.~Bonini, M.~Calandra, R.~Car, C.~Cavazzoni,
  D.~Ceresoli, G.~L. Chiarotti, M.~Cococcioni, I.~Dabo, A.~D. Corso,
  S.~de~Gironcoli, S.~Fabris, G.~Fratesi, R.~Gebauer, U.~Gerstmann,
  C.~Gougoussis, A.~Kokalj, M.~Lazzeri, L.~Martin-Samos, N.~Marzari, F.~Mauri,
  R.~Mazzarello, S.~Paolini, A.~Pasquarello, L.~Paulatto, C.~Sbraccia,
  S.~Scandolo, G.~Sclauzero, A.~P. Seitsonen, A.~Smogunov, P.~Umari, and R.~M.
  Wentzcovitch, {\it Quantum espresso: a modular and open-source software
  project for quantum simulations of materials},  {\em Journal of Physics:
  Condensed Matter} {\bf 21} (2009), no.~39 395502.

\bibitem{Hochberg:2016ajh}
Y.~Hochberg, T.~Lin, and K.~M. Zurek, {\it {Detecting Ultralight Bosonic Dark
  Matter via Absorption in Superconductors}},  {\em Phys. Rev.} {\bf D94}
  (2016), no.~1 015019, [\href{http://arxiv.org/abs/1604.06800}{
  arXiv:1604.06800}].

\bibitem{Hochberg:2016ntt}
Y.~Hochberg, Y.~Kahn, M.~Lisanti, C.~G. Tully, and K.~M. Zurek, {\it
  {Directional detection of dark matter with two-dimensional targets}},  {\em
  Phys. Lett.} {\bf B772} (2017) 239--246,
  [\href{http://arxiv.org/abs/1606.08849}{ arXiv:1606.08849}].

\bibitem{Hochberg:2016sqx}
Y.~Hochberg, T.~Lin, and K.~M. Zurek, {\it {Absorption of light dark matter in
  semiconductors}},  {\em Phys. Rev.} {\bf D95} (2017), no.~2 023013,
  [\href{http://arxiv.org/abs/1608.01994}{ arXiv:1608.01994}].

\bibitem{Hochberg:2017wce}
Y.~Hochberg, Y.~Kahn, M.~Lisanti, K.~M. Zurek, A.~G. Grushin, R.~Ilan, S.~M.
  Griffin, Z.-F. Liu, S.~F. Weber, and J.~B. Neaton, {\it {Detection of sub-MeV
  Dark Matter with Three-Dimensional Dirac Materials}},  {\em Phys. Rev.} {\bf
  D97} (2018), no.~1 015004, [\href{http://arxiv.org/abs/1708.08929}{
  arXiv:1708.08929}].

\bibitem{Cavoto:2017otc}
G.~Cavoto, F.~Luchetta, and A.~D. Polosa, {\it {Sub-GeV Dark Matter Detection
  with Electron Recoils in Carbon Nanotubes}},  {\em Phys. Lett.} {\bf B776}
  (2018) 338--344, [\href{http://arxiv.org/abs/1706.02487}{ arXiv:1706.02487}].

\bibitem{Knapen:2017ekk}
S.~Knapen, T.~Lin, M.~Pyle, and K.~M. Zurek, {\it {Detection of Light Dark
  Matter With Optical Phonons in Polar Materials}},
  \href{http://arxiv.org/abs/1712.06598}{ arXiv:1712.06598}.

\bibitem{Knapen:2017xzo}
S.~Knapen, T.~Lin, and K.~M. Zurek, {\it {Light Dark Matter: Models and
  Constraints}},  {\em Phys. Rev.} {\bf D96} (2017), no.~11 115021,
  [\href{http://arxiv.org/abs/1709.07882}{ arXiv:1709.07882}].

\bibitem{Griffin:2018bjn}
S.~Griffin, S.~Knapen, T.~Lin, and K.~M. Zurek, {\it {Directional Detection of
  Light Dark Matter with Polar Materials}},
  \href{http://arxiv.org/abs/1807.10291}{ arXiv:1807.10291}.

\bibitem{Liang:2018wte}
T.~Liang, B.~Zhu, R.~Ding, and T.~Li, {\it {Direct Detection of Axion-Like
  Particles in Bismuth-Based Topological Insulators}},
  \href{http://arxiv.org/abs/1807.11757}{ arXiv:1807.11757}.

\bibitem{PhysRevB.63.125108}
B.~Adolph, J.~Furthm\"uller, and F.~Bechstedt, {\it Optical properties of
  semiconductors using projector-augmented waves},  {\em Phys. Rev. B} {\bf 63}
  (Mar, 2001) 125108.

\bibitem{PhysRevB.58.4320}
P.~Delaney, B.~Kr\'alik, and S.~G. Louie, {\it Compton profiles of si:
  Pseudopotential calculation and reconstruction effects},  {\em Phys. Rev. B}
  {\bf 58} (Aug, 1998) 4320--4324.

\bibitem{MAKKONEN20051128}
I.~Makkonen, M.~Hakala, and M.~Puska, {\it Calculation of valence electron
  momentum densities using the projector augmented-wave method},  {\em Journal
  of Physics and Chemistry of Solids} {\bf 66} (2005), no.~6 1128 -- 1135.

\bibitem{PhysRevLett.43.1494}
D.~R. Hamann, M.~Schl\"uter, and C.~Chiang, {\it Norm-conserving
  pseudopotentials},  {\em Phys. Rev. Lett.} {\bf 43} (Nov, 1979) 1494--1497.

\bibitem{PhysRevB.41.7892}
D.~Vanderbilt, {\it Soft self-consistent pseudopotentials in a generalized
  eigenvalue formalism},  {\em Phys. Rev. B} {\bf 41} (Apr, 1990) 7892--7895.

\bibitem{PhysRevB.50.17953}
P.~E. Bl\"ochl, {\it Projector augmented-wave method},  {\em Phys. Rev. B} {\bf
  50} (Dec, 1994) 17953--17979.

\bibitem{PhysRevB.59.1758}
G.~Kresse and D.~Joubert, {\it From ultrasoft pseudopotentials to the projector
  augmented-wave method},  {\em Phys. Rev. B} {\bf 59} (Jan, 1999) 1758--1775.

\bibitem{HOLZWARTH2001329}
N.~Holzwarth, A.~Tackett, and G.~Matthews, {\it A projector augmented wave
  (paw) code for electronic structure calculations, part i: atompaw for
  generating atom-centered functions},  {\em Computer Physics Communications}
  {\bf 135} (2001), no.~3 329 -- 347.

\bibitem{PhysRevLett.77.3865}
J.~P. Perdew, K.~Burke, and M.~Ernzerhof, {\it Generalized gradient
  approximation made simple},  {\em Phys. Rev. Lett.} {\bf 77} (Oct, 1996)
  3865--3868.

\bibitem{GONZE20092582}
X.~Gonze, B.~Amadon, P.-M. Anglade, J.-M. Beuken, F.~Bottin, P.~Boulanger,
  F.~Bruneval, D.~Caliste, R.~Caracas, M.~Cote, T.~Deutsch, L.~Genovese,
  P.~Ghosez, M.~Giantomassi, S.~Goedecker, D.~Hamann, P.~Hermet, F.~Jollet,
  G.~Jomard, S.~Leroux, M.~Mancini, S.~Mazevet, M.~Oliveira, G.~Onida,
  Y.~Pouillon, T.~Rangel, G.-M. Rignanese, D.~Sangalli, R.~Shaltaf, M.~Torrent,
  M.~Verstraete, G.~Zerah, and J.~Zwanziger, {\it Abinit: First-principles
  approach to material and nanosystem properties},  {\em Computer Physics
  Communications} {\bf 180} (2009), no.~12 2582 -- 2615.

\end{thebibliography}\endgroup
\end{document}